\documentclass[prl,twocolumn,preprintnumbers,superscriptaddress,amsmath,amssymb]{revtex4-1}
\usepackage{graphicx}
\usepackage{subfigure}
\usepackage{mathrsfs}
\usepackage{amsfonts}
\usepackage{times}
\usepackage{amsmath}
\usepackage{leftidx}
\usepackage{tikz}
\usepackage{tikz-network}
\usepackage{color}
\usepackage[colorlinks,linkcolor=blue,citecolor=blue]{hyperref}
\usepackage[normalem]{ulem}

\newcommand{\ind}{\operatorname{ind}}
\newcommand{\Tr}{\operatorname{Tr}}

\usepackage{braket}
\usepackage{mathtools}

\def\be{\begin{equation}}
\def\ee{\end{equation}}

\newcommand{\vbkg}{{v^\ast}}

\begin{document}
\title{Coarse-grained Entanglement and Operator Growth in Anomalous Dynamics}
\author{Zongping Gong}
\affiliation{Max-Planck-Institut f\"ur Quantenoptik, Hans-Kopfermann-Stra{\ss}e 1, D-85748 Garching, Germany}
\author{Adam Nahum}
\affiliation{Theoretical  Physics,  University  of  Oxford,  Parks  Road,  Oxford  OX1  3PU,  United  Kingdom}
\affiliation{Laboratoire de Physique de l'\'Ecole Normale Sup\'erieure, CNRS, ENS \& Universit\'e PSL, Sorbonne Universit\'e, Universit\'e de Paris, 75005 Paris, France.}
\author{Lorenzo Piroli}
\affiliation{Max-Planck-Institut f\"ur Quantenoptik, Hans-Kopfermann-Stra{\ss}e 1, D-85748 Garching, Germany}
\affiliation{Philippe Meyer Institute, Physics Department, \'Ecole Normale Sup\'erieure (ENS), Universit\'e PSL, 24 rue Lhomond, F-75231 Paris, France}
\date{\today}

\begin{abstract}
	In two-dimensional Floquet systems, many-body localized dynamics in the bulk may give rise to a chaotic evolution at the one-dimensional edges that is characterized by a nonzero chiral topological index. Such \emph{anomalous} dynamics is qualitatively different from local-Hamiltonian evolution. Here we show how the presence of a nonzero index affects entanglement generation and the spreading of local operators, focusing on the coarse-grained description of generic systems. We tackle this problem by analyzing exactly solvable models of random quantum cellular automata (QCA) which generalize random circuits. We find that a nonzero index leads to asymmetric butterfly velocities with different diffusive broadening of the light cones, and to a modification of the order relations between the butterfly and entanglement velocities. We propose that these results can be understood via a generalization of the recently-introduced entanglement membrane theory, by allowing for a spacetime entropy current, which in the case of a generic QCA is fixed by the index. We work out the implications of this current on the entanglement ``membrane tension'' and show that the results for random QCA are recovered by identifying the topological index with a background velocity for the coarse-grained entanglement dynamics.
\end{abstract}

\maketitle

\begin{figure}
	\includegraphics[width=8.4cm, clip]{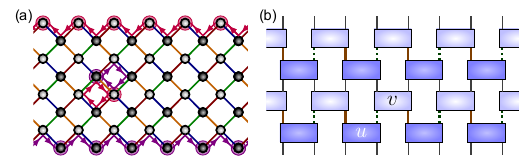}
	\caption{(a): Pictorial representation of LP evolution as the edge dynamics of a  Floquet qudit system. In the presence of an MBL phase in the bulk~\cite{footnoteMBL}, it is always possible to decompose the one-period Floquet operator as~\cite{po2016chiral} $U_{F}=U_{\mathrm{edge}} e^{-i H_{\mathrm{bulk}} T}$, where $H_{\mathrm{bulk}}$ is an MBL Hamiltonian, and $U_{\mathrm{edge}}$ is an effective $1$D evolution acting on qudits within a few localization lengths from the boundary. (b): Any LP evolution $\mathcal{U}(t)$ may be approximated by a QCA in Margolous form, i.e. the single time-step unitary operator $U$ admits a bilayer representation, where the local unitaries map a product of nearest-neighbor Hilbert spaces into another, with possibly different individual input or output dimensions.}
	\label{fig:fig1}
\end{figure}

\emph{Introduction.---} In many-body quantum systems subject to local-Hamiltonian dynamics, 
correlations propagate with a finite velocity, which determines an approximate causal cone~\cite{lieb1972finite}. This property, however,  holds beyond local-Hamiltonian evolution,  defining a more general class of \emph{locality-preserving} (LP) unitary dynamics that are termed quantum cellular automata (QCA) whenever the causal cone is strict~\cite{arrighi2019overview,farrelly2020review}. Models of LP evolution appear 
naturally at the boundary of certain Floquet systems displaying many-body localization (MBL) in the bulk~\cite{po2016chiral,po2017radical,Harper2017,duschatko2018tracking,fidkowski2019interacting} [cf. Fig.~\ref{fig:fig1}]. 

In one-dimension ($1$D), the mathematical theory of LP dynamics is well developed~\cite{arrighi2011unitarity,gross2012index,farrelly2014causal,cirac2017matrix,sahinoglu2018matrix,piroli2020quantum,piroli2021fermionic,gong2021topological}. 
A crucial result, proven in Ref.~\cite{gross2012index}, is that QCA are fully classified by a genuinely dynamical topological index. This result was recently generalized to include the more realistic case where the causal cone is only approximate~\cite{ranard2020converse}. 
Importantly, this topological index is zero if and only if the evolution is generated by a (quasi-)local Hamiltonian. Otherwise, the dynamics is said to be anomalous. As a natural application, this theory led to the discovery of new dynamical topological phases in $2$D Floquet MBL systems~\cite{po2016chiral,po2017radical,Harper2017,duschatko2018tracking,fidkowski2019interacting}, including the case of protecting symmetries~\cite{gong2020classification,zhang2021classification,gong2021topology}, which go beyond the cohomology paradigm \cite{Else2016,Potter2017,Roy2017}. 

Although the index was initially defined in terms of abstract operator algebras~\cite{gross2012index}, an equivalent definition, which  reflects an intuitive picture of quantum-information flow~\cite{duschatko2018tracking,gong2021topological,ranard2020converse}, was recently put forward. In turn, this made it possible to establish a lower bound on quantum scrambling in terms of the index, building a bridge between genuinely dynamical topological invariants and quantum chaos~\cite{gong2021topological}.

In this Letter, we develop a connection of  a different kind: while Ref.~\cite{gong2021topological} derived universal relations involving the index at the microscopic level, here we reveal its implications for the dynamics of generic systems at \emph{macroscopic} (hydrodynamic) scales. This is done within the framework of the entanglement membrane theory (EMT) developed in Refs.~\cite{nahum2017quantum,jonay2018coarse,zhou2020entanglement}. Following the logic of these works, where random unitary circuit (RUC) models played a key role, our approach is based on the analysis of random QCA, which we propose as minimal models for anomalous chaotic systems.

\emph{ Conventional EMT.---} Let us begin by recalling the basic aspects of the EMT~\cite{nahum2017quantum,jonay2018coarse}.
Throughout this work, we consider a 1D lattice of qudits associated with a Hilbert space $(\mathbb{C}^{d})^{\otimes 2L}$ ($2L$: system size) and a unitary dynamics dictated by the operator ${\cal U}(t):(\mathbb{C}^{d})^{\otimes 2L}\to(\mathbb{C}^{d})^{\otimes 2L}$, where time $t$ might be either continuous or discrete. 

\begin{figure}
	\includegraphics[width=8.4cm, clip]{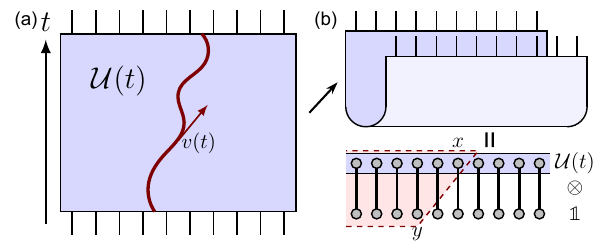}
	\caption{(a) Pictorial representation of a spacetime curve cutting through the unitary evolution operator ${\cal U}(t)$, in $(1+1)$D. For a given curve, the total line-tension is $\int s_{\mathrm{eq}} \mathcal{E}(v) d t$, where $v(t)$ is the local velocity. (b) The line tension may be obtained by viewing ${\cal U}(t)$ as a state, and computing the corresponding bipartite entanglement.}
	\label{fig:line_tension}
\end{figure}

The main object of the EMT is the so-called membrane tension 
(or line tension, in $1$D), which associates an entanglement cost with a given spacetime cut through the unitary operator ${\cal U}(t)$ [cf. Fig.~\ref{fig:line_tension}]. 
This quantity allows for an intuitive geometric picture for the coarse-grained entanglement dynamics.
The local tension $\mathcal{E}(v)$ is a function of the curve velocity ${v={\rm d}x/{\rm d}t}$, and the cost of a given curve is obtained by integrating $\mathcal{E}(v)$ along its length.
Then, the entanglement of a given interval $A$ in space at a given time is obtained by minimizing the integral of $\mathcal{E}(v)$ 
over all curves that separate a spacetime region that terminates on $A$ on the temporal boundary. As an example, we may consider the growth of the entanglement after a quench, for an infinite bipartite system with open boundary conditions: assuming homogeneous spacetime dynamics, we obtain
\begin{equation}\label{eq:local_production}
	S(x, t)= \min _{y}\left[t s_{\mathrm{eq}} \mathcal{E}\left(\frac{x-y}{t}\right)+S(y, 0)\right]\,,
\end{equation}
where $S(y, 0)$ is the entanglement of the initial state, while $s_{\mathrm{eq}}$ is the entanglement density reached at equilibrium~\cite{jonay2018coarse}. $S(x, t)$ here may indicate the von Neumann entanglement entropy, or (assuming the absence of conservation laws \cite{rakovszky2019sub,huang2020dynamics,rakovszky2019entanglement,zhou2020diffusive}) an arbitrary R\'enyi entropy~\cite{nielsen2002quantum}.
Holographic field theories give elegant examples of off-lattice systems where  $\mathcal{E}(v)$ is explicitly computable~\cite{mezei2018membrane,mezei2020exploring,agon2019bit}.

The EMT may be equivalently formulated in terms of a local entanglement production rate. In the bipartite setting above, the membrane picture is equivalent to a dynamical equation
$\frac{\partial S}{\partial t}=s_{\mathrm{eq}} \Gamma\left(\frac{\partial S}{\partial x}\right)$, where $\Gamma(s)$ is a local production rate dependent on the entanglement gradient~\cite{nahum2017quantum,jonay2018coarse}. Comparison with~\eqref{eq:local_production} reveals that $\Gamma(s)$ and $\mathcal{E}(v)$ are simply related by the Legendre transformation
\begin{equation}\label{eq:entanglement_production}
	\Gamma(s)=\min _{v}\left[\mathcal{E}(v)-\frac{v s}{s_{\mathrm{eq}}}\right]\,.
\end{equation}

The line tension also encodes information about the growth of local operators and, in general, must satisfy some basic constraints~\cite{jonay2018coarse}. First,  internal consistency of the coarse-grained picture requires that $\mathcal{E}(v)\geq 0$ and $\mathcal{E}^{''}(v)\geq 0$. Second, one may argue that the minimization in  Eq.~\eqref{eq:entanglement_production}  only involves membrane velocities within a range $[-v_{-}, v_{+}]$, where $v_\pm$ coincide with the left/right butterfly speeds $v_{L,R}$ that govern the growth of local operators. Although $v_{L}=v_R$ if spatial inversion symmetry is present, this is not generally true otherwise, even for local-Hamiltonian dynamics~\cite{liu2018asymmetric,stahl2018asymmetric,zhang2020asymmetric}. The minimum of $\mathcal{E}(v)$ is the \emph{entanglement velocity} $v_E$, quantifying the entanglement growth rate (rescaled by $s_{\rm eq}$) starting from a product state. Finally, one can show
\begin{equation}\label{eq:e_v_relation_ham}
\mathcal{E}(v_R)=v_R\,, \qquad \mathcal{E}(-v_L)=v_L\,,
\end{equation}
and ${\mathcal{E}^{\prime}(v_R)=-\mathcal{E}^{\prime}(-v_L)=1}$, implying ${v_{E} \leq \min \left\{v_{L}, v_{R}\right\}}$. This picture is  believed to hold for generic local-Hamiltonian and quantum-circuit evolution; our goal is to find whether and how it can be extended to anomalous dynamics. 

\emph{The Margolus form for QCA.---} As mentioned, the edge dynamics of the $2$D systems  in Fig.~\ref{fig:fig1}(a) is LP. This means that the single time-step unitary operator $U$ of the discrete evolution has the following property: for any local observable $O_j$ acting on site $j$,
the operator $U^\dagger O_j U$ is supported on a finite-neighborhood of $j$, up to exponentially decaying tails~\footnote{The index theory is in fact robust even when locality is preserved up to polynomially-decaying tails~\cite{ranard2020converse}. However, analogously to the case of local-Hamiltonian evolution, long-range tails might change some qualitative features of the dynamics, and are not considered in this work.}. 

Any LP dynamics may be approximated arbitrarily well by QCA, by ``chopping off'' the exponential tails~\cite{ranard2020converse}. In turn, it is known that any QCA may be expressed in the so-called Margolus form~\footnote{More precisely, any QCA may be represented in this way, up to grouping together finite sets of neighboring sites~\cite{farrelly2020review}} [cf. Fig.~\ref{fig:fig1}(b)], where $U$ is written as a two-layer product of two-site unitaries. This does not always define a quantum circuit because the dimensions of the local spaces associated with the ``virtual'' layer may differ from the physical ones: denoting by $p$, $q$ two integers such that ${d^2=pq}$,  we have
\be\label{eq:QCA_margolous}
U=(\otimes^L_{j=1} v_{2 j-1,2 j})(\otimes^L_{j=1}u_{2 j,2 j+1})
\ee where $u: \mathbb{C}^{d} \otimes \mathbb{C}^{d} \rightarrow \mathbb{C}^{p} \otimes \mathbb{C}^{q}$ and $v: \mathbb{C}^{q} \otimes \mathbb{C}^{p} \rightarrow \mathbb{C}^{d} \otimes \mathbb{C}^{d}$.  Given this representation, the topological index reads~\cite{gross2012index}
\begin{equation}
	{\rm ind}=\frac{1}{2} \ln \frac{q}{p}\,. 
\end{equation}
The unitary operation of translation by one site is a simple example with  ${{\rm ind}=\ln d}$. Note that for finite systems $\ind\neq0$ is only possible for periodic boundary conditions~\footnote{More precisely, a nonzero index is incompatible with a static spatial boundary that preserves unitarity. A  unitary boundary moving at coarse-grained velocity $\vbkg$ [defined in~\eqref{eq:identification}] can be constructed, as seen most easily when $U$ is the translation operator with $\vbkg=1$.}. For simplicity, we will always take $L\to\infty$, so that the boundary conditions become irrelevant.

The Margolus representation allows us to pinpoint the essential feature of anomalous dynamics which we have to take into account in order to generalize the conventional EMT. First, we note that the QCA in Fig.~\ref{fig:fig1}(b) can be viewed as a unitary tensor network (TN) and that, although the dimensions associated with given bonds may vary in space and time, unitarity requires that the input and output dimensions of each tensor must match. This gives rise to a non-trivial local conservation law, not accounted for in conventional EMT.

Physically, we can understand such local conservation law as a continuity equation of the form $\partial_\mu J^\mu=0$, in terms of a coarse-grained spacetime \emph{entropy current} $J^{\mu}$. For a unitary TN dynamics locally equilibrating to infinite temperature, $J^{\mu}$ has an explicit microscopic definition: Regarding the TN as a graph whose nodes are the unitaries and edges the bonds, we orient the latter in the direction of increasing time; then, along each bond we define the entropy current as a vector in the direction of its orientation, whose magnitude is equal to $\ln d_i$, with $d_i$ the associated local Hilbert-space dimension.

The coarse-grained spacetime entropy current is more general than the model above, where it can be introduced by ``counting'' microscopic bonds. For instance, it can be defined even when the equilibrium state is non-trivial and determined by the slow modes~\cite{InPrep}. In any case, it has important consequences on the properties of the membrane tension. In the following, we show how the EMT has to be modified in the presence of a non-trivial spacetime entropy current. The resulting \emph{generalized} EMT turns out to correctly capture the coarse-grained features of anomalous dynamics, finally revealing the hydrodynamic implications of the index.

\begin{figure}
	\includegraphics[scale=0.2]{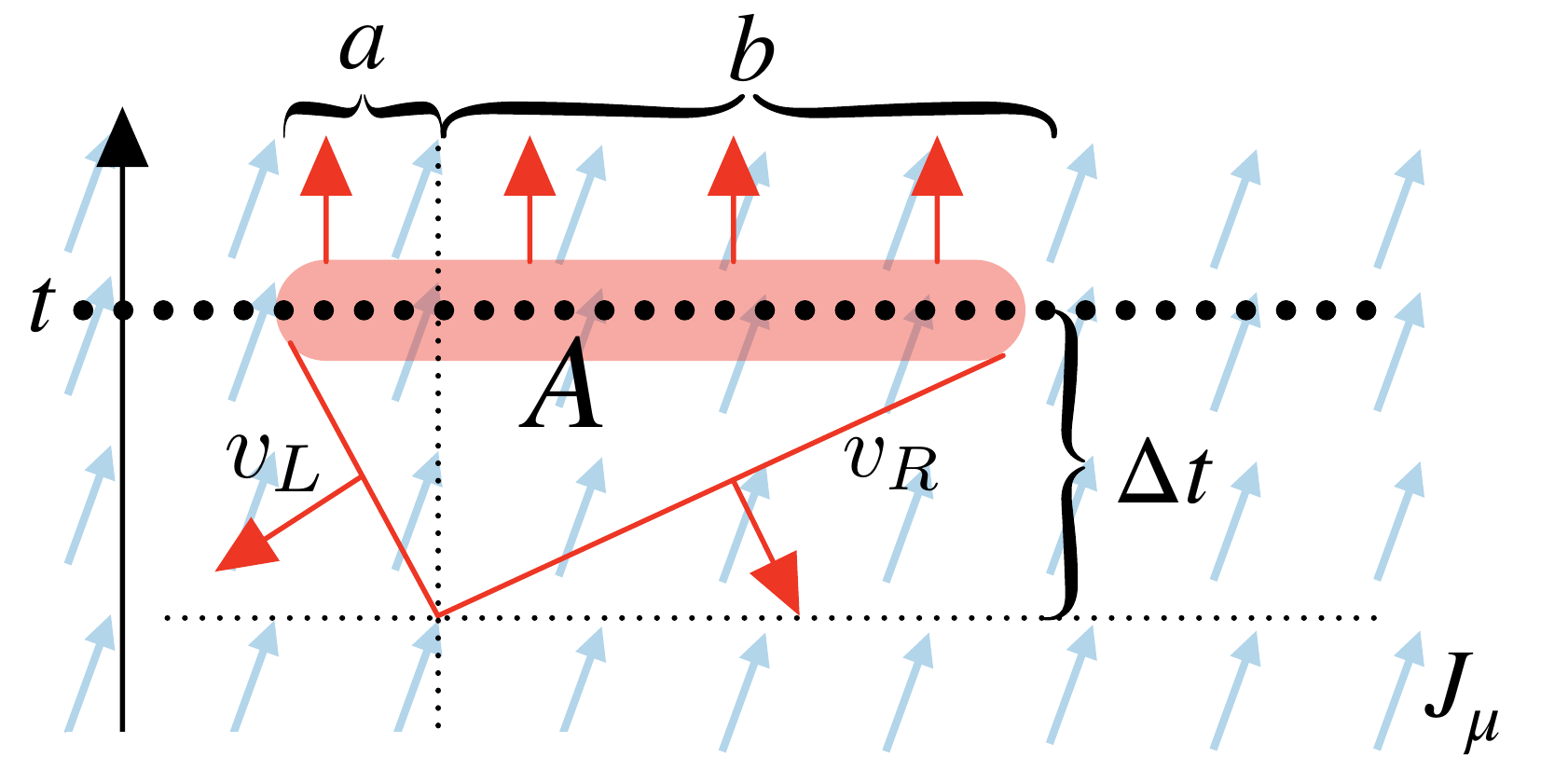}
	\caption{In the stationary regime, the divergence of the entropy current is vanishing, and the flux through any closed surface is zero. In order to compute the integral~\eqref{eq:def}, we choose a surface containing the optimal membranes and exploit the prescriptions of the EMT.}
	\label{fig:surface}
\end{figure}

\emph{Generalized EMT.---}   As a starting point to generalize the EMT, we assume there is a well-defined line-tension ${\cal E}(v)$ satisfying ${\cal E}(v), {\cal E}^{''}(v)\geq 0$ (as required for consistency of the hydrodynamic picture). We also postulate that there exists a spacetime entropy current $J^{\mu}(x,t)=(J_x,J_t)$ which governs the growth and transport of thermodynamic entropy. In particular, the density $s$ of thermodynamic entropy (we assume local equilibrium) is equal to $J_t$. 

Let us consider a stationary regime, focusing, for instance, on a finite interval $A$ at large times after a quench. Stationarity requires ${\partial_\mu J^\mu=0}$, and also implies that the thermodynamic entropy equals the von Neumann entropy, $S_A(t)$, i.e. 
\begin{equation}\label{eq:def}
	S_A(t)=\int_A {\rm d}x J_t(x)\,.
\end{equation}
Using the divergence theorem, the integral over $A$ may be obtained by integrating the current over any closed perimeter containing $A$ [cf. Fig.~\ref{fig:surface}]. In order to make contact with the EMT, we choose the perimeter to be a triangle whose bottom sides have slopes given by the butterfly velocities $-v_{L}$, $v_{R}$. A simple computation then yields $S_A(t)=(J_x\Delta t+J_ta)+(-J_x\Delta t+J_tb)$, where $\Delta t$, $a$ and $b$ are as in Fig.~\ref{fig:surface}. On the other hand, the sides of triangle minimize the line tension for the region $A$, as one can show by generalizing the arguments of~\cite{jonay2018coarse}. As a consequence $S_A(t)=s_\mathrm{eq}\left[\mathcal{E}(-v_L)\Delta t+\mathcal{E}(v_R) \Delta t\right]$. Identifying the individual terms coming from the two bottom sides of the triangle, and using $a/\Delta t=v_L$ and $b/\Delta t=v_R$, we find $J_t=\frac{s_\mathrm{eq}}{v_L+v_R}[\mathcal{E}(-v_L)+\mathcal{E}(v_R)]$ and
$J_x=\frac{s_\mathrm{eq}}{v_L+v_R}[{v_R}\mathcal{E}(-v_L)-{v_L}\mathcal{E}(v_R)]$. Defining now the \emph{background entropy velocity} ${\vbkg=J_x/J_t}$, and using $S_A(t)=s_{\rm eq} (a+b)$ (which follows from stationarity) we finally obtain
\be\label{eq:final}
\mathcal{E}(v_R)=v_R-\vbkg\,,\qquad \mathcal{E}(-v_L)=v_L+\vbkg\,.
\ee
This equation deviates from the conventional EMT, cf.~\eqref{eq:e_v_relation_ham}, and has important ramifications. In particular, combined with convexity, it implies
	$\mathcal{E}(v)\ge \left|v- v^\ast\right|$ 
and so $v_{E} \leq \min \left\{v_{R}-v^\ast, v_{L}+v^\ast\right\}$.

Using~\eqref{eq:final}, one can also argue that the relation between $\Gamma(s)$ and $\mathcal{E}(v)$ must be modified. Indeed, plugging~\eqref{eq:final} directly into~\eqref{eq:entanglement_production}, we see that $\Gamma(s)<0$ for some values of $s$. This is clearly an inconsistency, since $\Gamma(s)$ is the rate of entanglement growth. 
In order to guarantee positivity, one is led to the natural generalization
\begin{equation}
	\Gamma (s) = \min_v \left[\mathcal{E}(v) - \frac{vs}{s_{\rm eq}}\right] + v^\ast\frac{s}{s_{\rm eq}}\,.
	\label{eq:GE}
\end{equation}
From elementary properties of the Legendre transformation, we have the basic constraints $-s_{\mathrm{eq}} \Gamma^{\prime}\left(s_{\mathrm{eq}}\right)=v_{R}-v^*,
\quad s_{\mathrm{eq}} \Gamma^{\prime}\left(-s_{\mathrm{eq}}\right)=v_{L}+v^*
$. Note that by construction $\Gamma(0)=v_E$. Finally, differentiating  \eqref{eq:local_production} with respect to $t$, we obtain 
\begin{equation}\label{eq:entanglement_dynamics}
	\frac{\partial S}{\partial t} +v^\ast\frac{\partial S}{\partial x} = s_{\rm eq} \Gamma\left(\frac{\partial S}{\partial x}\right)\,.
\end{equation}
We see that the fundamental equation from the EMT, governing the coarse-grained entanglement dynamics, is modified by a constant velocity term. Importantly, $v^\ast$ is now left as a free parameter, and the conventional EMT is recovered for $v^{\ast}=0$. Note that without entropy production, i.e. $\Gamma\left(\frac{\partial S}{\partial x}\right)=0$, Eq.~\eqref{eq:entanglement_dynamics} still predicts a nonzero entropy change, which is qualitatively different from normal dynamics.

\emph{Models of random QCA.---}  In order to test the generalized EMT and identify the entropy-current velocity $v^\ast$,  we study concrete models of chaotic anomalous dynamics. We consider QCA of the form~\eqref{eq:QCA_margolous} where $u, v$ at different spacetime positions are drawn independently from the Haar random ensemble. Generalizing from the special case of RUCs, we expect the model also to capture universal aspects of random Floquet evolutions \cite{chan2018solution,chan2018spectral,bertini2018exact,sunderhauf2018localization,bertini2019entanglement,friedman2019spectral,chan2019eigenstate} and translationally invariant homogeneous systems~\cite{nahum2017quantum,jonay2018coarse,zhou2020entanglement,garratt2021local} if we restrict to the leading dynamics at large scales. We note that it is easy to construct explicit $2$D models with trivial bulk dynamics that display the random edge evolution considered here. This construction is detailed in the Supplemental Material (SM), where we also define even simpler random $1$D QCA which appear naturally in this context~\cite{SM}. 

As a first step, we analyze how the support of a localized traceless operator $O_0$ grows under the dynamics, which allows us to extract the butterfly velocities. We focus on the out-of-time-order correlator (OTOC) $\mathcal{C}(x, t) = {\rm tr}( \left[O_{0}, O^\prime_{2x}(t)\right]^{\dagger}\left[O_{0}, O^\prime_{2x}(t)\right])/2$,~\cite{larkin1969quasiclassical,kitaev2014talk,shenker2014black,shenker2014multiple,maldacena2016bound}. Here $O^\prime_{2x}(t)=\mathcal{U}(t)^\dag O^\prime_{2x}\mathcal{U}(t)$ and $O^\prime_{2x}$ is a traceless operator supported at site $2x$. Given the brickwork structure of the random QCA, the disorder-averaged OTOC $\overline{\mathcal{C}(x, t)}$ may be computed using the  approach developed for RUC, mapping the problem to the partition function of an Ising-like model~\cite{nahum2018operator,vonKeyserlingk2018operator}, see also~\cite{khemani2018operator,rakovszky2018diffusive,hunter2018operator,hunter2019unitary,zhou2019emergent,bertini2020scrambling}. Additional technical complications arise due to the ``staggered'' structure of the dynamics, alternating physical and virtual Hilbert spaces [cf. Fig.~\ref{fig:fig1}(b)]. Nevertheless, a fully analytic expression may be obtained~\cite{SM}, and in the hydrodynamic limit of large spacetime scales, it simply reads
$\overline{{\cal C} (x,t)}\simeq \Phi(\frac{v_{L} t+x}{\sigma_L}) \Phi(\frac{v_{R} t-x}{\sigma_R})$. Here $\Phi(y)=(2 \pi)^{-1/2} \int_{-\infty}^{y} e^{-x^{2} / 2} \mathrm{~d} x$ and
\begin{equation}
	\begin{split}
		&\;\;\;\;\;\;\;\;\;\;\;\;\;\;\;\;\;v_{L}=\frac{p^{2} q^{2}-q(p+q)+1}{p^{2} q^{2}-1}, \\
		&\sigma_L=\frac{\sqrt{t} \sqrt{q \left[p^3 q^2+p^2 q \left(q^2-3\right)+p-q^3+q\right]}}{p^2q^2-1}\,,
	\end{split}
	\label{vLsL}	
\end{equation}
while $v_R$ and $\sigma_R$ are obtained by exchanging $p \leftrightarrow q$. We see that the coarse-grained OTOC has the same form as RUC~\cite{nahum2018operator,vonKeyserlingk2018operator}, being characterized by two propagating fronts inside which $\overline{{\cal C} (x,t)}\simeq 1$, i.e. information is fully scrambled. However, the fronts propagate with asymmetric butterfly velocities and widths, with the faster front being the narrowest one (in contrast to the quantum circuit models constructed in Ref.~\cite{stahl2018asymmetric}). Note that when ${\rm ind}=0$, i.e. $p=q$, we recover the result for RUC~\cite{nahum2018operator,vonKeyserlingk2018operator}. 
In the other limit where ${\rm ind}=\ln d$, i.e.  $p=1$ and  $q=d^2$,  the random QCA consists of a two-step evolution in which a right translation is followed by a layer of random unitary gates. Since the shift does not increase the operator support, $\sigma_{L,R}$ in \eqref{vLsL} are then simply those of a RUC evolved up to a time $t^\prime=t/2$. 

\begin{figure}
	\begin{center}
		\includegraphics[scale=0.33]{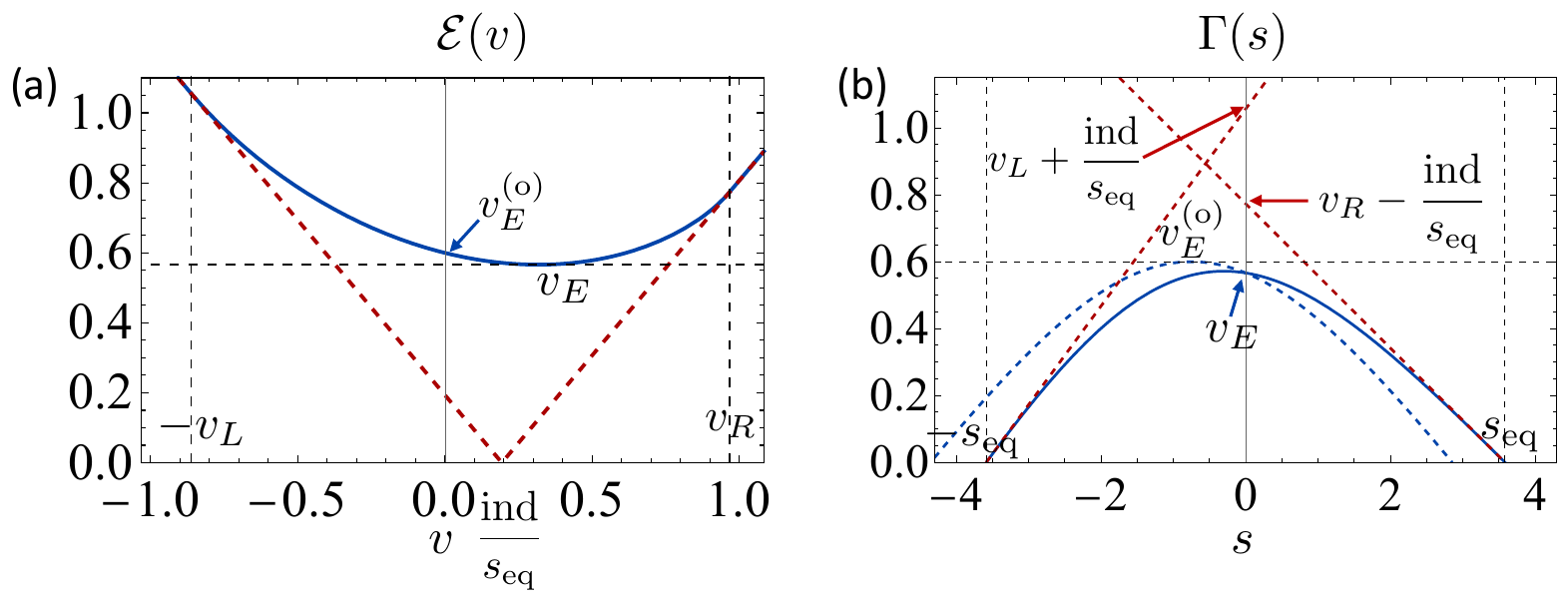}
	\end{center}
	\caption{(a) Entanglement line tension $\mathcal{E}(v)$ and (b) production $\Gamma(s)$ (in terms of R\'enyi-2 entropy) for the random QCA model with $d=6$ and $p=3$, $q=12$. The red dashed lines refer to the tangents at (a) $(v_{L,R},\mathcal{E}(\mp v_{L,R})=v_{L,R}\pm \ind/s_{\rm eq})$; (b) $(\pm s_{\rm eq},0)$. The blue dashed line in (b) corresponds to $\Gamma(s)-(\ind/s_{\rm eq})(s/s_{\rm eq})$, whose maximum gives $v^{(\rm o)}_E$.}
	\label{fig:membrane}
\end{figure}

\emph{Entanglement dynamics.---} Next, we move on to compute the line-tension, cf. Fig.~\ref{fig:line_tension}. Formally, we introduce a doubled Hilbert space $\mathcal{H}=(\mathbb{C}^{d})^{\otimes 2L}\otimes (\mathbb{C}^{d})^{\otimes 2L}$ along with the maximally entangled state $\ket{I}=\ket{\phi_+}^{\otimes 2L}$, where $\ket{\phi_+}\equiv\sum^d_{n=1}\ket{n}\otimes\ket{n}/\sqrt{d}$ and $n$ runs over a basis of $\mathbb{C}^d$. 
This allows us to vectorize the evolution operator as $\ket{{\cal U}(t)}=({\cal U}(t)\otimes \openone)\ket{I}\in {\cal H}$, with the two Hilbert spaces associated with its input and output degrees of freedom. Considering a bipartition of the system with boundary at site $x$ and $y$ in the input and output, respectively, we define the operator entanglement~\cite{zanardi2001entanglement,prosen2007operator,dubail2017entanglement,zhou2017operator} $S^{(\mathrm{o})}(y-x, t)$ as the associated entanglement entropy of $\ket{{\cal U}(t)}$. The line-tension may then be computed via~\cite{jonay2018coarse} $\mathcal{E}(v)=\lim _{t \to \infty} S^{(\mathrm{o})}(v t, t)/(s_{\rm eq}t)$, for $v$ in the range ${(-v_-, v_+)}$. Our random QCA has no conserved quantity and our unit length increment ${{\rm d}x=1}$ is defined to contain two sites, so that  ${s_{\rm eq}=\ln d^2=\ln (p q)}$. $v^{(\rm o)}=\mathcal{E}(0)$ is the \emph{operator-entanglement velocity}. 

Averaged R\'enyi-$n$ entropies are hard to compute, but the mapping to an Ising partition function gives access to the averaged purity and its logarithm $-\ln \overline{e^{-S^{(\mathrm{o})}_2(x,t)}}$. Since this average is taken ``inside the logarithm'', it differs from the averaged R\'enyi-$2$ entropy $\overline{S^{(\mathrm{o})}_2}$. 
However, the former is sufficient to see the key relationships obeyed by $\mathcal{E}$. 
The line tension for $\overline{S^{(\mathrm{o})}_2}$ can be understood as a perturbatively ``dressed'' version of that for $-\ln \overline{e^{-S^{(\mathrm{o})}_2(x,t)}}$ \cite{zhou2020entanglement}: the difference between the two vanishes as ${d\rightarrow \infty}$, and in fact is numerically small even for finite~$d$~\footnote{For RUC, it was proven that  the two quantities coincide up to order $1/(d^8 \ln d)$~\cite{zhou2019emergent}.}. So we approximate {$t^{-1} {\overline{S^{(\mathrm{o})}_2(x,t)}} \simeq - t^{-1}\ln \overline{e^{-S^{(\mathrm{o})}_2(x,t)}}$},  yielding~\cite{SM}
\begin{widetext}
	\begin{equation}
		\mathcal{E}_{2}(v)\!=\!\log_{d^2} \! \frac{(p q+1)^{2}(p q-1)}{p q \sqrt{\left(p^{2}-1\right)\left(q^{2}-1\right)}}\!-v \log_{d^2}\! \frac{\sqrt{\left(1-r_{p}\right)^{2} v^{2}+4 r_{p}}-\left(1+r_{p}\right) v}{2(1+v)}-\log_{d^2} \! \frac{1+r_{p}+\sqrt{\left(1-r_{p}\right)^{2} v^{2}+4 r_{p}}}{\left(1-v^{2}\right) \sqrt{r_{p}}}\,,
		\label{E2v}
	\end{equation}
\end{widetext}
where $r_{p}=p(q^{2}-1)/[q(p^{2}-1)]$. As a main difference from the case of RUC, the minimum is at a nonzero velocity ${v_{m}=(r_{p}-1)/[2(r_{p}+1)]}$ [cf. Fig.~\ref{fig:membrane}(a)]. 
The  value ${\cal E}_2(v_m)$ yields  the entanglement speed $v_E=\log_{d^2}  [(p q+1)^{2}/(2 \sqrt{p q}(p+q))]$, as confirmed by directly computing
the growth of state R\'enyi-$2$ entropy following a quench from a product state 
\cite{SM}.  We stress that $v_{L,R}$ and $v_E$ 
do not depend on ${\rm ind}$ in a universal way. This could be expected from the study of RUC, where asymmetric butterfly velocities might be realized by specific arrangements of the local unitaries~\cite{stahl2018asymmetric}.

Crucially, we see that ${\cal E}^{\prime}_2(\pm v_\pm)=\pm1$, where $v_{\pm}=v_{R,L}$, and that Eq.~\eqref{eq:final} is satisfied, after the identification
\be
v^\ast=\frac{\ind}{s_{\mathrm{eq}}}\,.
\label{eq:identification}
\ee
This is our final main result: it states that the index, a microscopic dynamical topological invariant, appears at the hydrodynamic level as a constant background velocity for the coarse-grained entanglement dynamics. Based on this identification, the index also determines the qualitative features of the rate $\Gamma(s)$, which is shown in Fig.~\ref{fig:membrane}. Further details on the random QCA, including a computation of the so-called tripartite mutual information~\cite{hosur2016chaos} and its relation to the index, are reported in the SM~\cite{SM}.

\emph{Outlook.}--- Our results open up several possibilities for future research. First, when viewing anomalous 1D dynamics as boundaries of 2D Floquet systems, it would be interesting to investigate the corrections to our theory when the assumption of ideal localization in the bulk is relaxed. In this case, we expect subleading effects emerging, due to a slow entropy flow from the boundary to the bulk, and vice versa.
It would also be natural to apply our picture based on spacetime entropy currents to more general situations with inhomogeneous backgrounds, as models with genuinely spacetime-dependent entropy currents may be constructed by introducing additional structure. Next, it would be interesting to explore how chaotic anomalous dynamics is modified by local conservation laws, such as $U(1)$ charges as done for RUC \cite{rakovszky2018diffusive,khemani2018operator}. Studies along this direction could reveal an intriguing effect of the index on the otherwise purely diffusive behavior of the charge. Finally, two natural generalizations of our study include adding
randomized measurements~\cite{yaodong2018quantum,skinner2019measurement,chan2019unitary,fan2020self,choi2020quantum,gullans2020dynamical,ippoliti2021entanglement,jian2020measurement} and higher dimensions, where the theory of QCA is much more open~\cite{haah2018nontrivial,haah2019clifford,freedman2019group,freedman2020classification}.

\emph{Acknowledgments.---} 
We acknowledge Ignacio Cirac and David Huse for helpful discussions. We thank the anonymous Referee for valuable comments on the manuscript. Z.G. is supported by the Max-Planck-Harvard Research Center for Quantum Optics (MPHQ).

\bibliography{./bibliography}

\begin{thebibliography}{87}%
\makeatletter
\providecommand \@ifxundefined [1]{%
 \@ifx{#1\undefined}
}%
\providecommand \@ifnum [1]{%
 \ifnum #1\expandafter \@firstoftwo
 \else \expandafter \@secondoftwo
 \fi
}%
\providecommand \@ifx [1]{%
 \ifx #1\expandafter \@firstoftwo
 \else \expandafter \@secondoftwo
 \fi
}%
\providecommand \natexlab [1]{#1}%
\providecommand \enquote  [1]{``#1''}%
\providecommand \bibnamefont  [1]{#1}%
\providecommand \bibfnamefont [1]{#1}%
\providecommand \citenamefont [1]{#1}%
\providecommand \href@noop [0]{\@secondoftwo}%
\providecommand \href [0]{\begingroup \@sanitize@url \@href}%
\providecommand \@href[1]{\@@startlink{#1}\@@href}%
\providecommand \@@href[1]{\endgroup#1\@@endlink}%
\providecommand \@sanitize@url [0]{\catcode `\\12\catcode `\$12\catcode
  `\&12\catcode `\#12\catcode `\^12\catcode `\_12\catcode `\%12\relax}%
\providecommand \@@startlink[1]{}%
\providecommand \@@endlink[0]{}%
\providecommand \url  [0]{\begingroup\@sanitize@url \@url }%
\providecommand \@url [1]{\endgroup\@href {#1}{\urlprefix }}%
\providecommand \urlprefix  [0]{URL }%
\providecommand \Eprint [0]{\href }%
\providecommand \doibase [0]{http://dx.doi.org/}%
\providecommand \selectlanguage [0]{\@gobble}%
\providecommand \bibinfo  [0]{\@secondoftwo}%
\providecommand \bibfield  [0]{\@secondoftwo}%
\providecommand \translation [1]{[#1]}%
\providecommand \BibitemOpen [0]{}%
\providecommand \bibitemStop [0]{}%
\providecommand \bibitemNoStop [0]{.\EOS\space}%
\providecommand \EOS [0]{\spacefactor3000\relax}%
\providecommand \BibitemShut  [1]{\csname bibitem#1\endcsname}%
\let\auto@bib@innerbib\@empty
\bibitem [{\citenamefont {Lieb}\ and\ \citenamefont
  {Robinson}(1972)}]{lieb1972finite}%
  \BibitemOpen
  \bibfield  {author} {\bibinfo {author} {\bibfnamefont {E.}~\bibnamefont
  {Lieb}}\ and\ \bibinfo {author} {\bibfnamefont {D.}~\bibnamefont
  {Robinson}},\ }\href {\doibase 10.1007/BF01645779} {\bibfield  {journal}
  {\bibinfo  {journal} {Comm. Math. Phys.}\ }\textbf {\bibinfo {volume} {28}},\
  \bibinfo {pages} {251} (\bibinfo {year} {1972})}\BibitemShut {NoStop}%
\bibitem [{\citenamefont {Arrighi}(2019)}]{arrighi2019overview}%
  \BibitemOpen
  \bibfield  {author} {\bibinfo {author} {\bibfnamefont {P.}~\bibnamefont
  {Arrighi}},\ }\href {\doibase 10.1007/s11047-019-09762-6} {\bibfield
  {journal} {\bibinfo  {journal} {Natural Comp.}\ }\textbf {\bibinfo {volume}
  {18}},\ \bibinfo {pages} {885} (\bibinfo {year} {2019})}\BibitemShut
  {NoStop}%
\bibitem [{\citenamefont {Farrelly}(2020)}]{farrelly2020review}%
  \BibitemOpen
  \bibfield  {author} {\bibinfo {author} {\bibfnamefont {T.}~\bibnamefont
  {Farrelly}},\ }\href {\doibase 10.22331/q-2020-11-30-368} {\bibfield
  {journal} {\bibinfo  {journal} {Quantum}\ }\textbf {\bibinfo {volume} {4}},\
  \bibinfo {pages} {368} (\bibinfo {year} {2020})}\BibitemShut {NoStop}%
\bibitem [{\citenamefont {Po}\ \emph {et~al.}(2016)\citenamefont {Po},
  \citenamefont {Fidkowski}, \citenamefont {Morimoto}, \citenamefont {Potter},\
  and\ \citenamefont {Vishwanath}}]{po2016chiral}%
  \BibitemOpen
  \bibfield  {author} {\bibinfo {author} {\bibfnamefont {H.~C.}\ \bibnamefont
  {Po}}, \bibinfo {author} {\bibfnamefont {L.}~\bibnamefont {Fidkowski}},
  \bibinfo {author} {\bibfnamefont {T.}~\bibnamefont {Morimoto}}, \bibinfo
  {author} {\bibfnamefont {A.~C.}\ \bibnamefont {Potter}}, \ and\ \bibinfo
  {author} {\bibfnamefont {A.}~\bibnamefont {Vishwanath}},\ }\href {\doibase
  10.1103/PhysRevX.6.041070} {\bibfield  {journal} {\bibinfo  {journal} {Phys.
  Rev. X}\ }\textbf {\bibinfo {volume} {6}},\ \bibinfo {pages} {041070}
  (\bibinfo {year} {2016})}\BibitemShut {NoStop}%
\bibitem [{\citenamefont {Po}\ \emph {et~al.}(2017)\citenamefont {Po},
  \citenamefont {Fidkowski}, \citenamefont {Vishwanath},\ and\ \citenamefont
  {Potter}}]{po2017radical}%
  \BibitemOpen
  \bibfield  {author} {\bibinfo {author} {\bibfnamefont {H.~C.}\ \bibnamefont
  {Po}}, \bibinfo {author} {\bibfnamefont {L.}~\bibnamefont {Fidkowski}},
  \bibinfo {author} {\bibfnamefont {A.}~\bibnamefont {Vishwanath}}, \ and\
  \bibinfo {author} {\bibfnamefont {A.~C.}\ \bibnamefont {Potter}},\ }\href
  {\doibase 10.1103/PhysRevB.96.245116} {\bibfield  {journal} {\bibinfo
  {journal} {Phys. Rev. B}\ }\textbf {\bibinfo {volume} {96}},\ \bibinfo
  {pages} {245116} (\bibinfo {year} {2017})}\BibitemShut {NoStop}%
\bibitem [{\citenamefont {Harper}\ and\ \citenamefont
  {Roy}(2017)}]{Harper2017}%
  \BibitemOpen
  \bibfield  {author} {\bibinfo {author} {\bibfnamefont {F.}~\bibnamefont
  {Harper}}\ and\ \bibinfo {author} {\bibfnamefont {R.}~\bibnamefont {Roy}},\
  }\href {\doibase 10.1103/PhysRevLett.118.115301} {\bibfield  {journal}
  {\bibinfo  {journal} {Phys. Rev. Lett.}\ }\textbf {\bibinfo {volume} {118}},\
  \bibinfo {pages} {115301} (\bibinfo {year} {2017})}\BibitemShut {NoStop}%
\bibitem [{\citenamefont {Duschatko}\ \emph {et~al.}(2018)\citenamefont
  {Duschatko}, \citenamefont {Dumitrescu},\ and\ \citenamefont
  {Potter}}]{duschatko2018tracking}%
  \BibitemOpen
  \bibfield  {author} {\bibinfo {author} {\bibfnamefont {B.~R.}\ \bibnamefont
  {Duschatko}}, \bibinfo {author} {\bibfnamefont {P.~T.}\ \bibnamefont
  {Dumitrescu}}, \ and\ \bibinfo {author} {\bibfnamefont {A.~C.}\ \bibnamefont
  {Potter}},\ }\href {\doibase 10.1103/PhysRevB.98.054309} {\bibfield
  {journal} {\bibinfo  {journal} {Phys. Rev. B}\ }\textbf {\bibinfo {volume}
  {98}},\ \bibinfo {pages} {054309} (\bibinfo {year} {2018})}\BibitemShut
  {NoStop}%
\bibitem [{\citenamefont {Fidkowski}\ \emph {et~al.}(2019)\citenamefont
  {Fidkowski}, \citenamefont {Po}, \citenamefont {Potter},\ and\ \citenamefont
  {Vishwanath}}]{fidkowski2019interacting}%
  \BibitemOpen
  \bibfield  {author} {\bibinfo {author} {\bibfnamefont {L.}~\bibnamefont
  {Fidkowski}}, \bibinfo {author} {\bibfnamefont {H.~C.}\ \bibnamefont {Po}},
  \bibinfo {author} {\bibfnamefont {A.~C.}\ \bibnamefont {Potter}}, \ and\
  \bibinfo {author} {\bibfnamefont {A.}~\bibnamefont {Vishwanath}},\ }\href
  {\doibase 10.1103/PhysRevB.99.085115} {\bibfield  {journal} {\bibinfo
  {journal} {Phys. Rev. B}\ }\textbf {\bibinfo {volume} {99}},\ \bibinfo
  {pages} {085115} (\bibinfo {year} {2019})}\BibitemShut {NoStop}%
\bibitem [{\citenamefont {Arrighi}\ \emph {et~al.}(2011)\citenamefont
  {Arrighi}, \citenamefont {Nesme},\ and\ \citenamefont
  {Werner}}]{arrighi2011unitarity}%
  \BibitemOpen
  \bibfield  {author} {\bibinfo {author} {\bibfnamefont {P.}~\bibnamefont
  {Arrighi}}, \bibinfo {author} {\bibfnamefont {V.}~\bibnamefont {Nesme}}, \
  and\ \bibinfo {author} {\bibfnamefont {R.}~\bibnamefont {Werner}},\ }\href
  {\doibase 10.1016/j.jcss.2010.05.004} {\bibfield  {journal} {\bibinfo
  {journal} {J. Comp. Syst. Sciences}\ }\textbf {\bibinfo {volume} {77}},\
  \bibinfo {pages} {372} (\bibinfo {year} {2011})}\BibitemShut {NoStop}%
\bibitem [{\citenamefont {Gross}\ \emph {et~al.}(2012)\citenamefont {Gross},
  \citenamefont {Nesme}, \citenamefont {Vogts},\ and\ \citenamefont
  {Werner}}]{gross2012index}%
  \BibitemOpen
  \bibfield  {author} {\bibinfo {author} {\bibfnamefont {D.}~\bibnamefont
  {Gross}}, \bibinfo {author} {\bibfnamefont {V.}~\bibnamefont {Nesme}},
  \bibinfo {author} {\bibfnamefont {H.}~\bibnamefont {Vogts}}, \ and\ \bibinfo
  {author} {\bibfnamefont {R.~F.}\ \bibnamefont {Werner}},\ }\href {\doibase
  10.1007/s00220-012-1423-1} {\bibfield  {journal} {\bibinfo  {journal} {Comm.
  Math. Phys.}\ }\textbf {\bibinfo {volume} {310}},\ \bibinfo {pages} {419}
  (\bibinfo {year} {2012})}\BibitemShut {NoStop}%
\bibitem [{\citenamefont {Farrelly}\ and\ \citenamefont
  {Short}(2014)}]{farrelly2014causal}%
  \BibitemOpen
  \bibfield  {author} {\bibinfo {author} {\bibfnamefont {T.~C.}\ \bibnamefont
  {Farrelly}}\ and\ \bibinfo {author} {\bibfnamefont {A.~J.}\ \bibnamefont
  {Short}},\ }\href {\doibase 10.1103/PhysRevA.89.012302} {\bibfield  {journal}
  {\bibinfo  {journal} {Phys. Rev. A}\ }\textbf {\bibinfo {volume} {89}},\
  \bibinfo {pages} {012302} (\bibinfo {year} {2014})}\BibitemShut {NoStop}%
\bibitem [{\citenamefont {Cirac}\ \emph {et~al.}(2017)\citenamefont {Cirac},
  \citenamefont {Perez-Garcia}, \citenamefont {Schuch},\ and\ \citenamefont
  {Verstraete}}]{cirac2017matrix}%
  \BibitemOpen
  \bibfield  {author} {\bibinfo {author} {\bibfnamefont {J.~I.}\ \bibnamefont
  {Cirac}}, \bibinfo {author} {\bibfnamefont {D.}~\bibnamefont {Perez-Garcia}},
  \bibinfo {author} {\bibfnamefont {N.}~\bibnamefont {Schuch}}, \ and\ \bibinfo
  {author} {\bibfnamefont {F.}~\bibnamefont {Verstraete}},\ }\href {\doibase
  10.1088/1742-5468/aa7e55} {\bibfield  {journal} {\bibinfo  {journal} {J.
  Stat. Mech.}\ }\textbf {\bibinfo {volume} {2017}},\ \bibinfo {pages} {083105}
  (\bibinfo {year} {2017})}\BibitemShut {NoStop}%
\bibitem [{\citenamefont {Sahinoglu}\ \emph {et~al.}(2018)\citenamefont
  {Sahinoglu}, \citenamefont {Shukla}, \citenamefont {Bi},\ and\ \citenamefont
  {Chen}}]{sahinoglu2018matrix}%
  \BibitemOpen
  \bibfield  {author} {\bibinfo {author} {\bibfnamefont {M.~B.}\ \bibnamefont
  {Sahinoglu}}, \bibinfo {author} {\bibfnamefont {S.~K.}\ \bibnamefont
  {Shukla}}, \bibinfo {author} {\bibfnamefont {F.}~\bibnamefont {Bi}}, \ and\
  \bibinfo {author} {\bibfnamefont {X.}~\bibnamefont {Chen}},\ }\href {\doibase
  10.1103/PhysRevB.98.245122} {\bibfield  {journal} {\bibinfo  {journal} {Phys.
  Rev. B}\ }\textbf {\bibinfo {volume} {98}},\ \bibinfo {pages} {245122}
  (\bibinfo {year} {2018})}\BibitemShut {NoStop}%
\bibitem [{\citenamefont {Piroli}\ and\ \citenamefont
  {Cirac}(2020)}]{piroli2020quantum}%
  \BibitemOpen
  \bibfield  {author} {\bibinfo {author} {\bibfnamefont {L.}~\bibnamefont
  {Piroli}}\ and\ \bibinfo {author} {\bibfnamefont {J.~I.}\ \bibnamefont
  {Cirac}},\ }\href {\doibase 10.1103/PhysRevLett.125.190402} {\bibfield
  {journal} {\bibinfo  {journal} {Phys. Rev. Lett.}\ }\textbf {\bibinfo
  {volume} {125}},\ \bibinfo {pages} {190402} (\bibinfo {year}
  {2020})}\BibitemShut {NoStop}%
\bibitem [{\citenamefont {Piroli}\ \emph {et~al.}(2021)\citenamefont {Piroli},
  \citenamefont {Turzillo}, \citenamefont {Shukla},\ and\ \citenamefont
  {Cirac}}]{piroli2021fermionic}%
  \BibitemOpen
  \bibfield  {author} {\bibinfo {author} {\bibfnamefont {L.}~\bibnamefont
  {Piroli}}, \bibinfo {author} {\bibfnamefont {A.}~\bibnamefont {Turzillo}},
  \bibinfo {author} {\bibfnamefont {S.~K.}\ \bibnamefont {Shukla}}, \ and\
  \bibinfo {author} {\bibfnamefont {J.~I.}\ \bibnamefont {Cirac}},\ }\href
  {\doibase 10.1088/1742-5468/abd30f} {\bibfield  {journal} {\bibinfo
  {journal} {J. Stat. Mech.}\ }\textbf {\bibinfo {volume} {2021}},\ \bibinfo
  {pages} {013107} (\bibinfo {year} {2021})}\BibitemShut {NoStop}%
\bibitem [{\citenamefont {Gong}\ \emph {et~al.}(2021)\citenamefont {Gong},
  \citenamefont {Piroli},\ and\ \citenamefont {Cirac}}]{gong2021topological}%
  \BibitemOpen
  \bibfield  {author} {\bibinfo {author} {\bibfnamefont {Z.}~\bibnamefont
  {Gong}}, \bibinfo {author} {\bibfnamefont {L.}~\bibnamefont {Piroli}}, \ and\
  \bibinfo {author} {\bibfnamefont {J.~I.}\ \bibnamefont {Cirac}},\ }\href
  {\doibase 10.1103/PhysRevLett.126.160601} {\bibfield  {journal} {\bibinfo
  {journal} {Phys. Rev. Lett.}\ }\textbf {\bibinfo {volume} {126}},\ \bibinfo
  {pages} {160601} (\bibinfo {year} {2021})}\BibitemShut {NoStop}%
\bibitem [{\citenamefont {Ranard}\ \emph {et~al.}(2020)\citenamefont {Ranard},
  \citenamefont {Walter},\ and\ \citenamefont
  {Witteveen}}]{ranard2020converse}%
  \BibitemOpen
  \bibfield  {author} {\bibinfo {author} {\bibfnamefont {D.}~\bibnamefont
  {Ranard}}, \bibinfo {author} {\bibfnamefont {M.}~\bibnamefont {Walter}}, \
  and\ \bibinfo {author} {\bibfnamefont {F.}~\bibnamefont {Witteveen}},\ }\href
  {https://arxiv.org/abs/2012.00741} {\bibfield  {journal} {\bibinfo  {journal}
  {arXiv:2012.00741}\ } (\bibinfo {year} {2020})}\BibitemShut {NoStop}%
\bibitem [{\citenamefont {Gong}\ \emph {et~al.}(2020)\citenamefont {Gong},
  \citenamefont {S\"underhauf}, \citenamefont {Schuch},\ and\ \citenamefont
  {Cirac}}]{gong2020classification}%
  \BibitemOpen
  \bibfield  {author} {\bibinfo {author} {\bibfnamefont {Z.}~\bibnamefont
  {Gong}}, \bibinfo {author} {\bibfnamefont {C.}~\bibnamefont {S\"underhauf}},
  \bibinfo {author} {\bibfnamefont {N.}~\bibnamefont {Schuch}}, \ and\ \bibinfo
  {author} {\bibfnamefont {J.~I.}\ \bibnamefont {Cirac}},\ }\href {\doibase
  10.1103/PhysRevLett.124.100402} {\bibfield  {journal} {\bibinfo  {journal}
  {Phys. Rev. Lett.}\ }\textbf {\bibinfo {volume} {124}},\ \bibinfo {pages}
  {100402} (\bibinfo {year} {2020})}\BibitemShut {NoStop}%
\bibitem [{\citenamefont {Zhang}\ and\ \citenamefont
  {Levin}(2021)}]{zhang2021classification}%
  \BibitemOpen
  \bibfield  {author} {\bibinfo {author} {\bibfnamefont {C.}~\bibnamefont
  {Zhang}}\ and\ \bibinfo {author} {\bibfnamefont {M.}~\bibnamefont {Levin}},\
  }\href {\doibase 10.1103/PhysRevB.103.064302} {\bibfield  {journal} {\bibinfo
   {journal} {Phys. Rev. B}\ }\textbf {\bibinfo {volume} {103}},\ \bibinfo
  {pages} {064302} (\bibinfo {year} {2021})}\BibitemShut {NoStop}%
\bibitem [{\citenamefont {Gong}\ and\ \citenamefont
  {Guaita}(2021)}]{gong2021topology}%
  \BibitemOpen
  \bibfield  {author} {\bibinfo {author} {\bibfnamefont {Z.}~\bibnamefont
  {Gong}}\ and\ \bibinfo {author} {\bibfnamefont {T.}~\bibnamefont {Guaita}},\
  }\href {https://arxiv.org/abs/2106.05044} {\bibfield  {journal} {\bibinfo
  {journal} {arXiv:2106.05044}\ } (\bibinfo {year} {2021})}\BibitemShut
  {NoStop}%
\bibitem [{\citenamefont {Else}\ and\ \citenamefont {Nayak}(2016)}]{Else2016}%
  \BibitemOpen
  \bibfield  {author} {\bibinfo {author} {\bibfnamefont {D.~V.}\ \bibnamefont
  {Else}}\ and\ \bibinfo {author} {\bibfnamefont {C.}~\bibnamefont {Nayak}},\
  }\href {\doibase 10.1103/PhysRevB.93.201103} {\bibfield  {journal} {\bibinfo
  {journal} {Phys. Rev. B}\ }\textbf {\bibinfo {volume} {93}},\ \bibinfo
  {pages} {201103} (\bibinfo {year} {2016})}\BibitemShut {NoStop}%
\bibitem [{\citenamefont {Potter}\ and\ \citenamefont
  {Morimoto}(2017)}]{Potter2017}%
  \BibitemOpen
  \bibfield  {author} {\bibinfo {author} {\bibfnamefont {A.~C.}\ \bibnamefont
  {Potter}}\ and\ \bibinfo {author} {\bibfnamefont {T.}~\bibnamefont
  {Morimoto}},\ }\href {\doibase 10.1103/PhysRevB.95.155126} {\bibfield
  {journal} {\bibinfo  {journal} {Phys. Rev. B}\ }\textbf {\bibinfo {volume}
  {95}},\ \bibinfo {pages} {155126} (\bibinfo {year} {2017})}\BibitemShut
  {NoStop}%
\bibitem [{\citenamefont {Roy}\ and\ \citenamefont {Harper}(2017)}]{Roy2017}%
  \BibitemOpen
  \bibfield  {author} {\bibinfo {author} {\bibfnamefont {R.}~\bibnamefont
  {Roy}}\ and\ \bibinfo {author} {\bibfnamefont {F.}~\bibnamefont {Harper}},\
  }\href {\doibase 10.1103/PhysRevB.95.195128} {\bibfield  {journal} {\bibinfo
  {journal} {Phys. Rev. B}\ }\textbf {\bibinfo {volume} {95}},\ \bibinfo
  {pages} {195128} (\bibinfo {year} {2017})}\BibitemShut {NoStop}%
\bibitem [{\citenamefont {Nahum}\ \emph {et~al.}(2017)\citenamefont {Nahum},
  \citenamefont {Ruhman}, \citenamefont {Vijay},\ and\ \citenamefont
  {Haah}}]{nahum2017quantum}%
  \BibitemOpen
  \bibfield  {author} {\bibinfo {author} {\bibfnamefont {A.}~\bibnamefont
  {Nahum}}, \bibinfo {author} {\bibfnamefont {J.}~\bibnamefont {Ruhman}},
  \bibinfo {author} {\bibfnamefont {S.}~\bibnamefont {Vijay}}, \ and\ \bibinfo
  {author} {\bibfnamefont {J.}~\bibnamefont {Haah}},\ }\href {\doibase
  10.1103/PhysRevX.7.031016} {\bibfield  {journal} {\bibinfo  {journal} {Phys.
  Rev. X}\ }\textbf {\bibinfo {volume} {7}},\ \bibinfo {pages} {031016}
  (\bibinfo {year} {2017})}\BibitemShut {NoStop}%
\bibitem [{\citenamefont {Jonay}\ \emph {et~al.}(2018)\citenamefont {Jonay},
  \citenamefont {Huse},\ and\ \citenamefont {Nahum}}]{jonay2018coarse}%
  \BibitemOpen
  \bibfield  {author} {\bibinfo {author} {\bibfnamefont {C.}~\bibnamefont
  {Jonay}}, \bibinfo {author} {\bibfnamefont {D.~A.}\ \bibnamefont {Huse}}, \
  and\ \bibinfo {author} {\bibfnamefont {A.}~\bibnamefont {Nahum}},\ }\href
  {https://arxiv.org/abs/1803.00089} {\bibfield  {journal} {\bibinfo  {journal}
  {arXiv:1803.00089}\ } (\bibinfo {year} {2018})}\BibitemShut {NoStop}%
\bibitem [{\citenamefont {Zhou}\ and\ \citenamefont
  {Nahum}(2020)}]{zhou2020entanglement}%
  \BibitemOpen
  \bibfield  {author} {\bibinfo {author} {\bibfnamefont {T.}~\bibnamefont
  {Zhou}}\ and\ \bibinfo {author} {\bibfnamefont {A.}~\bibnamefont {Nahum}},\
  }\href {\doibase 10.1103/PhysRevX.10.031066} {\bibfield  {journal} {\bibinfo
  {journal} {Phys. Rev. X}\ }\textbf {\bibinfo {volume} {10}},\ \bibinfo
  {pages} {031066} (\bibinfo {year} {2020})}\BibitemShut {NoStop}%
\bibitem [{foo()}]{footnoteMBL}%
  \BibitemOpen
  \href@noop {} {}\bibinfo {note} {To be more precise, MBL is believed to be
  unstable in two and higher dimensions (i.e., strict MBL requires fine-tuning
  of the dynamics). However, localization can persist for times
  double-exponentially long in disorder strength, so for practical purposes can
  resemble a phase (see Ref.~\cite{po2016chiral} for discussion in the QCA
  context).}\BibitemShut {Stop}%
\bibitem [{\citenamefont {Rakovszky}\ \emph
  {et~al.}(2019{\natexlab{a}})\citenamefont {Rakovszky}, \citenamefont
  {Pollmann},\ and\ \citenamefont {Von~Keyserlingk}}]{rakovszky2019sub}%
  \BibitemOpen
  \bibfield  {author} {\bibinfo {author} {\bibfnamefont {T.}~\bibnamefont
  {Rakovszky}}, \bibinfo {author} {\bibfnamefont {F.}~\bibnamefont {Pollmann}},
  \ and\ \bibinfo {author} {\bibfnamefont {C.}~\bibnamefont
  {Von~Keyserlingk}},\ }\href {\doibase 10.1103/PhysRevLett.122.250602}
  {\bibfield  {journal} {\bibinfo  {journal} {Phys. Rev. Lett.}\ }\textbf
  {\bibinfo {volume} {122}},\ \bibinfo {pages} {250602} (\bibinfo {year}
  {2019}{\natexlab{a}})}\BibitemShut {NoStop}%
\bibitem [{\citenamefont {Huang}(2020)}]{huang2020dynamics}%
  \BibitemOpen
  \bibfield  {author} {\bibinfo {author} {\bibfnamefont {Y.}~\bibnamefont
  {Huang}},\ }\href {\doibase 10.1088/2633-1357/abd1e2} {\bibfield  {journal}
  {\bibinfo  {journal} {IOP SciNotes}\ }\textbf {\bibinfo {volume} {1}},\
  \bibinfo {pages} {035205} (\bibinfo {year} {2020})}\BibitemShut {NoStop}%
\bibitem [{\citenamefont {Rakovszky}\ \emph
  {et~al.}(2019{\natexlab{b}})\citenamefont {Rakovszky}, \citenamefont {von
  Keyserlingk},\ and\ \citenamefont {Pollmann}}]{rakovszky2019entanglement}%
  \BibitemOpen
  \bibfield  {author} {\bibinfo {author} {\bibfnamefont {T.}~\bibnamefont
  {Rakovszky}}, \bibinfo {author} {\bibfnamefont {C.}~\bibnamefont {von
  Keyserlingk}}, \ and\ \bibinfo {author} {\bibfnamefont {F.}~\bibnamefont
  {Pollmann}},\ }\href {\doibase 10.1103/PhysRevB.100.125139} {\bibfield
  {journal} {\bibinfo  {journal} {Phys. Rev. B}\ }\textbf {\bibinfo {volume}
  {100}},\ \bibinfo {pages} {125139} (\bibinfo {year}
  {2019}{\natexlab{b}})}\BibitemShut {NoStop}%
\bibitem [{\citenamefont {Zhou}\ and\ \citenamefont
  {Ludwig}(2020)}]{zhou2020diffusive}%
  \BibitemOpen
  \bibfield  {author} {\bibinfo {author} {\bibfnamefont {T.}~\bibnamefont
  {Zhou}}\ and\ \bibinfo {author} {\bibfnamefont {A.~W.}\ \bibnamefont
  {Ludwig}},\ }\href {\doibase 10.1103/PhysRevResearch.2.033020} {\bibfield
  {journal} {\bibinfo  {journal} {Phys. Rev. Research}\ }\textbf {\bibinfo
  {volume} {2}},\ \bibinfo {pages} {033020} (\bibinfo {year}
  {2020})}\BibitemShut {NoStop}%
\bibitem [{\citenamefont {Nielsen}\ and\ \citenamefont
  {Chuang}(2002)}]{nielsen2002quantum}%
  \BibitemOpen
  \bibfield  {author} {\bibinfo {author} {\bibfnamefont {M.~A.}\ \bibnamefont
  {Nielsen}}\ and\ \bibinfo {author} {\bibfnamefont {I.}~\bibnamefont
  {Chuang}},\ }\href@noop {} {\emph {\bibinfo {title} {Quantum computation and
  quantum information}}}\ (\bibinfo  {publisher} {Cambridge University Press},\
  \bibinfo {year} {2002})\BibitemShut {NoStop}%
\bibitem [{\citenamefont {Mezei}(2018)}]{mezei2018membrane}%
  \BibitemOpen
  \bibfield  {author} {\bibinfo {author} {\bibfnamefont {M.}~\bibnamefont
  {Mezei}},\ }\href@noop {} {\bibfield  {journal} {\bibinfo  {journal} {Phys.
  Rev. D}\ }\textbf {\bibinfo {volume} {98}},\ \bibinfo {pages} {106025}
  (\bibinfo {year} {2018})}\BibitemShut {NoStop}%
\bibitem [{\citenamefont {Mezei}\ and\ \citenamefont
  {Virrueta}(2020)}]{mezei2020exploring}%
  \BibitemOpen
  \bibfield  {author} {\bibinfo {author} {\bibfnamefont {M.}~\bibnamefont
  {Mezei}}\ and\ \bibinfo {author} {\bibfnamefont {J.}~\bibnamefont
  {Virrueta}},\ }\href {\doibase 10.1007/JHEP02(2020)013} {\bibfield  {journal}
  {\bibinfo  {journal} {JHEP}\ }\textbf {\bibinfo {volume} {2020}},\ \bibinfo
  {pages} {1} (\bibinfo {year} {2020})}\BibitemShut {NoStop}%
\bibitem [{\citenamefont {Ag{\'o}n}\ and\ \citenamefont
  {Mezei}(2019)}]{agon2019bit}%
  \BibitemOpen
  \bibfield  {author} {\bibinfo {author} {\bibfnamefont {C.~A.}\ \bibnamefont
  {Ag{\'o}n}}\ and\ \bibinfo {author} {\bibfnamefont {M.}~\bibnamefont
  {Mezei}},\ }\href {https://arxiv.org/abs/1910.12909} {\bibfield  {journal}
  {\bibinfo  {journal} {arXiv:1910.12909}\ } (\bibinfo {year}
  {2019})}\BibitemShut {NoStop}%
\bibitem [{\citenamefont {Liu}\ \emph {et~al.}(2018)\citenamefont {Liu},
  \citenamefont {Garrison}, \citenamefont {Deng}, \citenamefont {Gong},\ and\
  \citenamefont {Gorshkov}}]{liu2018asymmetric}%
  \BibitemOpen
  \bibfield  {author} {\bibinfo {author} {\bibfnamefont {F.}~\bibnamefont
  {Liu}}, \bibinfo {author} {\bibfnamefont {J.~R.}\ \bibnamefont {Garrison}},
  \bibinfo {author} {\bibfnamefont {D.-L.}\ \bibnamefont {Deng}}, \bibinfo
  {author} {\bibfnamefont {Z.-X.}\ \bibnamefont {Gong}}, \ and\ \bibinfo
  {author} {\bibfnamefont {A.~V.}\ \bibnamefont {Gorshkov}},\ }\href {\doibase
  10.1103/PhysRevLett.121.250404} {\bibfield  {journal} {\bibinfo  {journal}
  {Phys. Rev. Lett.}\ }\textbf {\bibinfo {volume} {121}},\ \bibinfo {pages}
  {250404} (\bibinfo {year} {2018})}\BibitemShut {NoStop}%
\bibitem [{\citenamefont {Stahl}\ \emph {et~al.}(2018)\citenamefont {Stahl},
  \citenamefont {Khemani},\ and\ \citenamefont {Huse}}]{stahl2018asymmetric}%
  \BibitemOpen
  \bibfield  {author} {\bibinfo {author} {\bibfnamefont {C.}~\bibnamefont
  {Stahl}}, \bibinfo {author} {\bibfnamefont {V.}~\bibnamefont {Khemani}}, \
  and\ \bibinfo {author} {\bibfnamefont {D.~A.}\ \bibnamefont {Huse}},\ }\href
  {https://arxiv.org/abs/1812.05589} {\bibfield  {journal} {\bibinfo  {journal}
  {arXiv:1812.05589}\ } (\bibinfo {year} {2018})}\BibitemShut {NoStop}%
\bibitem [{\citenamefont {Zhang}\ and\ \citenamefont
  {Khemani}(2020)}]{zhang2020asymmetric}%
  \BibitemOpen
  \bibfield  {author} {\bibinfo {author} {\bibfnamefont {Y.-L.}\ \bibnamefont
  {Zhang}}\ and\ \bibinfo {author} {\bibfnamefont {V.}~\bibnamefont
  {Khemani}},\ }\href {\doibase 10.21468/SciPostPhys.9.2.024} {\bibfield
  {journal} {\bibinfo  {journal} {SciPost Phys.}\ }\textbf {\bibinfo {volume}
  {9}},\ \bibinfo {pages} {24} (\bibinfo {year} {2020})}\BibitemShut {NoStop}%
\bibitem [{Note1()}]{Note1}%
  \BibitemOpen
  \bibinfo {note} {The index theory is in fact robust even when locality is
  preserved up to polynomially-decaying tails~\cite {ranard2020converse}.
  However, analogously to the case of local-Hamiltonian evolution, long-range
  tails might change some qualitative features of the dynamics, and are not
  considered in this work.}\BibitemShut {Stop}%
\bibitem [{Note2()}]{Note2}%
  \BibitemOpen
  \bibinfo {note} {More precisely, any QCA may be represented in this way, up
  to grouping together finite sets of neighboring sites~\cite
  {farrelly2020review}}\BibitemShut {NoStop}%
\bibitem [{Note3()}]{Note3}%
  \BibitemOpen
  \bibinfo {note} {More precisely, a nonzero index is incompatible with a
  static spatial boundary that preserves unitarity. A unitary boundary moving
  at coarse-grained velocity ${v^\ast }$ [defined in~\protect \textup {\hbox
  {\mathsurround \z@ \protect \normalfont (\ignorespaces \ref
  {eq:identification}\unskip \@@italiccorr )}}] can be constructed, as seen
  most easily when $U$ is the translation operator with ${v^\ast
  }=1$.}\BibitemShut {Stop}%
\bibitem [{InP()}]{InPrep}%
  \BibitemOpen
  \href@noop {} {}\bibinfo {note} {Details will be provided in future
  works.}\BibitemShut {Stop}%
\bibitem [{\citenamefont {Chan}\ \emph
  {et~al.}(2018{\natexlab{a}})\citenamefont {Chan}, \citenamefont {De~Luca},\
  and\ \citenamefont {Chalker}}]{chan2018solution}%
  \BibitemOpen
  \bibfield  {author} {\bibinfo {author} {\bibfnamefont {A.}~\bibnamefont
  {Chan}}, \bibinfo {author} {\bibfnamefont {A.}~\bibnamefont {De~Luca}}, \
  and\ \bibinfo {author} {\bibfnamefont {J.~T.}\ \bibnamefont {Chalker}},\
  }\href {\doibase 10.1103/PhysRevX.8.041019} {\bibfield  {journal} {\bibinfo
  {journal} {Phys. Rev. X}\ }\textbf {\bibinfo {volume} {8}},\ \bibinfo {pages}
  {041019} (\bibinfo {year} {2018}{\natexlab{a}})}\BibitemShut {NoStop}%
\bibitem [{\citenamefont {Chan}\ \emph
  {et~al.}(2018{\natexlab{b}})\citenamefont {Chan}, \citenamefont {De~Luca},\
  and\ \citenamefont {Chalker}}]{chan2018spectral}%
  \BibitemOpen
  \bibfield  {author} {\bibinfo {author} {\bibfnamefont {A.}~\bibnamefont
  {Chan}}, \bibinfo {author} {\bibfnamefont {A.}~\bibnamefont {De~Luca}}, \
  and\ \bibinfo {author} {\bibfnamefont {J.~T.}\ \bibnamefont {Chalker}},\
  }\href {\doibase 10.1103/PhysRevLett.121.060601} {\bibfield  {journal}
  {\bibinfo  {journal} {Phys. Rev. Lett.}\ }\textbf {\bibinfo {volume} {121}},\
  \bibinfo {pages} {060601} (\bibinfo {year} {2018}{\natexlab{b}})}\BibitemShut
  {NoStop}%
\bibitem [{\citenamefont {Bertini}\ \emph {et~al.}(2018)\citenamefont
  {Bertini}, \citenamefont {Kos},\ and\ \citenamefont
  {Prosen}}]{bertini2018exact}%
  \BibitemOpen
  \bibfield  {author} {\bibinfo {author} {\bibfnamefont {B.}~\bibnamefont
  {Bertini}}, \bibinfo {author} {\bibfnamefont {P.}~\bibnamefont {Kos}}, \ and\
  \bibinfo {author} {\bibfnamefont {T.}~\bibnamefont {Prosen}},\ }\href
  {\doibase 10.1103/PhysRevLett.121.264101} {\bibfield  {journal} {\bibinfo
  {journal} {Phys. Rev. Lett.}\ }\textbf {\bibinfo {volume} {121}},\ \bibinfo
  {pages} {264101} (\bibinfo {year} {2018})}\BibitemShut {NoStop}%
\bibitem [{\citenamefont {S\"underhauf}\ \emph {et~al.}(2018)\citenamefont
  {S\"underhauf}, \citenamefont {P\'erez-Garc\'{\i}a}, \citenamefont {Huse},
  \citenamefont {Schuch},\ and\ \citenamefont
  {Cirac}}]{sunderhauf2018localization}%
  \BibitemOpen
  \bibfield  {author} {\bibinfo {author} {\bibfnamefont {C.}~\bibnamefont
  {S\"underhauf}}, \bibinfo {author} {\bibfnamefont {D.}~\bibnamefont
  {P\'erez-Garc\'{\i}a}}, \bibinfo {author} {\bibfnamefont {D.~A.}\
  \bibnamefont {Huse}}, \bibinfo {author} {\bibfnamefont {N.}~\bibnamefont
  {Schuch}}, \ and\ \bibinfo {author} {\bibfnamefont {J.~I.}\ \bibnamefont
  {Cirac}},\ }\href {\doibase 10.1103/PhysRevB.98.134204} {\bibfield  {journal}
  {\bibinfo  {journal} {Phys. Rev. B}\ }\textbf {\bibinfo {volume} {98}},\
  \bibinfo {pages} {134204} (\bibinfo {year} {2018})}\BibitemShut {NoStop}%
\bibitem [{\citenamefont {Bertini}\ \emph {et~al.}(2019)\citenamefont
  {Bertini}, \citenamefont {Kos},\ and\ \citenamefont
  {Prosen}}]{bertini2019entanglement}%
  \BibitemOpen
  \bibfield  {author} {\bibinfo {author} {\bibfnamefont {B.}~\bibnamefont
  {Bertini}}, \bibinfo {author} {\bibfnamefont {P.}~\bibnamefont {Kos}}, \ and\
  \bibinfo {author} {\bibfnamefont {T.}~\bibnamefont {Prosen}},\ }\href
  {\doibase 10.1103/PhysRevX.9.021033} {\bibfield  {journal} {\bibinfo
  {journal} {Phys. Rev. X}\ }\textbf {\bibinfo {volume} {9}},\ \bibinfo {pages}
  {021033} (\bibinfo {year} {2019})}\BibitemShut {NoStop}%
\bibitem [{\citenamefont {Friedman}\ \emph {et~al.}(2019)\citenamefont
  {Friedman}, \citenamefont {Chan}, \citenamefont {De~Luca},\ and\
  \citenamefont {Chalker}}]{friedman2019spectral}%
  \BibitemOpen
  \bibfield  {author} {\bibinfo {author} {\bibfnamefont {A.~J.}\ \bibnamefont
  {Friedman}}, \bibinfo {author} {\bibfnamefont {A.}~\bibnamefont {Chan}},
  \bibinfo {author} {\bibfnamefont {A.}~\bibnamefont {De~Luca}}, \ and\
  \bibinfo {author} {\bibfnamefont {J.~T.}\ \bibnamefont {Chalker}},\ }\href
  {\doibase 10.1103/PhysRevLett.123.210603} {\bibfield  {journal} {\bibinfo
  {journal} {Phys. Rev. Lett.}\ }\textbf {\bibinfo {volume} {123}},\ \bibinfo
  {pages} {210603} (\bibinfo {year} {2019})}\BibitemShut {NoStop}%
\bibitem [{\citenamefont {Chan}\ \emph
  {et~al.}(2019{\natexlab{a}})\citenamefont {Chan}, \citenamefont {De~Luca},\
  and\ \citenamefont {Chalker}}]{chan2019eigenstate}%
  \BibitemOpen
  \bibfield  {author} {\bibinfo {author} {\bibfnamefont {A.}~\bibnamefont
  {Chan}}, \bibinfo {author} {\bibfnamefont {A.}~\bibnamefont {De~Luca}}, \
  and\ \bibinfo {author} {\bibfnamefont {J.~T.}\ \bibnamefont {Chalker}},\
  }\href {\doibase 10.1103/PhysRevLett.122.220601} {\bibfield  {journal}
  {\bibinfo  {journal} {Phys. Rev. Lett.}\ }\textbf {\bibinfo {volume} {122}},\
  \bibinfo {pages} {220601} (\bibinfo {year} {2019}{\natexlab{a}})}\BibitemShut
  {NoStop}%
\bibitem [{\citenamefont {Garratt}\ and\ \citenamefont
  {Chalker}(2021)}]{garratt2021local}%
  \BibitemOpen
  \bibfield  {author} {\bibinfo {author} {\bibfnamefont {S.~J.}\ \bibnamefont
  {Garratt}}\ and\ \bibinfo {author} {\bibfnamefont {J.~T.}\ \bibnamefont
  {Chalker}},\ }\href {\doibase 10.1103/PhysRevX.11.021051} {\bibfield
  {journal} {\bibinfo  {journal} {Phys. Rev. X}\ }\textbf {\bibinfo {volume}
  {11}},\ \bibinfo {pages} {021051} (\bibinfo {year} {2021})}\BibitemShut
  {NoStop}%
\bibitem [{SM()}]{SM}%
  \BibitemOpen
  \href@noop {} {}\bibinfo {note} {See Supplemental Material, which includes
  Refs.~\cite{paris2013asymptotics,paris2013asymptoticsII,sunderhauf2019quantum,schnaack2019tripartite,kudler2021information},
  for further details.}\BibitemShut {Stop}%
\bibitem [{\citenamefont {Larkin}\ and\ \citenamefont
  {Ovchinnikov}(1969)}]{larkin1969quasiclassical}%
  \BibitemOpen
  \bibfield  {author} {\bibinfo {author} {\bibfnamefont {A.}~\bibnamefont
  {Larkin}}\ and\ \bibinfo {author} {\bibfnamefont {Y.~N.}\ \bibnamefont
  {Ovchinnikov}},\ }\href@noop {} {\bibfield  {journal} {\bibinfo  {journal}
  {Sov Phys JETP}\ }\textbf {\bibinfo {volume} {28}},\ \bibinfo {pages} {1200}
  (\bibinfo {year} {1969})}\BibitemShut {NoStop}%
\bibitem [{\citenamefont {Kitaev}()}]{kitaev2014talk}%
  \BibitemOpen
  \bibfield  {author} {\bibinfo {author} {\bibfnamefont {A.}~\bibnamefont
  {Kitaev}},\ }\href@noop {} {\enquote {\bibinfo {title} {Hidden correlations
  in the hawking radiation and thermal noise},}\ }\bibinfo {note} {In The
  Fundamental Physics Prize Symposium (2014)}\BibitemShut {NoStop}%
\bibitem [{\citenamefont {Shenker}\ and\ \citenamefont
  {Stanford}(2014{\natexlab{a}})}]{shenker2014black}%
  \BibitemOpen
  \bibfield  {author} {\bibinfo {author} {\bibfnamefont {S.~H.}\ \bibnamefont
  {Shenker}}\ and\ \bibinfo {author} {\bibfnamefont {D.}~\bibnamefont
  {Stanford}},\ }\href {\doibase 10.1007/JHEP03(2014)067} {\bibfield  {journal}
  {\bibinfo  {journal} {JHEP}\ }\textbf {\bibinfo {volume} {2014}},\ \bibinfo
  {pages} {67} (\bibinfo {year} {2014}{\natexlab{a}})}\BibitemShut {NoStop}%
\bibitem [{\citenamefont {Shenker}\ and\ \citenamefont
  {Stanford}(2014{\natexlab{b}})}]{shenker2014multiple}%
  \BibitemOpen
  \bibfield  {author} {\bibinfo {author} {\bibfnamefont {S.~H.}\ \bibnamefont
  {Shenker}}\ and\ \bibinfo {author} {\bibfnamefont {D.}~\bibnamefont
  {Stanford}},\ }\href {\doibase 10.1007/JHEP03(2014)067} {\bibfield  {journal}
  {\bibinfo  {journal} {JHEP}\ }\textbf {\bibinfo {volume} {2014}},\ \bibinfo
  {pages} {46} (\bibinfo {year} {2014}{\natexlab{b}})}\BibitemShut {NoStop}%
\bibitem [{\citenamefont {Maldacena}\ \emph {et~al.}(2016)\citenamefont
  {Maldacena}, \citenamefont {Shenker},\ and\ \citenamefont
  {Stanford}}]{maldacena2016bound}%
  \BibitemOpen
  \bibfield  {author} {\bibinfo {author} {\bibfnamefont {J.}~\bibnamefont
  {Maldacena}}, \bibinfo {author} {\bibfnamefont {S.~H.}\ \bibnamefont
  {Shenker}}, \ and\ \bibinfo {author} {\bibfnamefont {D.}~\bibnamefont
  {Stanford}},\ }\href {\doibase 10.1007/JHEP08(2016)106} {\bibfield  {journal}
  {\bibinfo  {journal} {JHEP}\ }\textbf {\bibinfo {volume} {2016}},\ \bibinfo
  {pages} {106} (\bibinfo {year} {2016})}\BibitemShut {NoStop}%
\bibitem [{\citenamefont {Nahum}\ \emph {et~al.}(2018)\citenamefont {Nahum},
  \citenamefont {Vijay},\ and\ \citenamefont {Haah}}]{nahum2018operator}%
  \BibitemOpen
  \bibfield  {author} {\bibinfo {author} {\bibfnamefont {A.}~\bibnamefont
  {Nahum}}, \bibinfo {author} {\bibfnamefont {S.}~\bibnamefont {Vijay}}, \ and\
  \bibinfo {author} {\bibfnamefont {J.}~\bibnamefont {Haah}},\ }\href {\doibase
  10.1103/PhysRevX.8.021014} {\bibfield  {journal} {\bibinfo  {journal} {Phys.
  Rev. X}\ }\textbf {\bibinfo {volume} {8}},\ \bibinfo {pages} {021014}
  (\bibinfo {year} {2018})}\BibitemShut {NoStop}%
\bibitem [{\citenamefont {von Keyserlingk}\ \emph {et~al.}(2018)\citenamefont
  {von Keyserlingk}, \citenamefont {Rakovszky}, \citenamefont {Pollmann},\ and\
  \citenamefont {Sondhi}}]{vonKeyserlingk2018operator}%
  \BibitemOpen
  \bibfield  {author} {\bibinfo {author} {\bibfnamefont {C.~W.}\ \bibnamefont
  {von Keyserlingk}}, \bibinfo {author} {\bibfnamefont {T.}~\bibnamefont
  {Rakovszky}}, \bibinfo {author} {\bibfnamefont {F.}~\bibnamefont {Pollmann}},
  \ and\ \bibinfo {author} {\bibfnamefont {S.~L.}\ \bibnamefont {Sondhi}},\
  }\href {\doibase 10.1103/PhysRevX.8.021013} {\bibfield  {journal} {\bibinfo
  {journal} {Phys. Rev. X}\ }\textbf {\bibinfo {volume} {8}},\ \bibinfo {pages}
  {021013} (\bibinfo {year} {2018})}\BibitemShut {NoStop}%
\bibitem [{\citenamefont {Khemani}\ \emph {et~al.}(2018)\citenamefont
  {Khemani}, \citenamefont {Vishwanath},\ and\ \citenamefont
  {Huse}}]{khemani2018operator}%
  \BibitemOpen
  \bibfield  {author} {\bibinfo {author} {\bibfnamefont {V.}~\bibnamefont
  {Khemani}}, \bibinfo {author} {\bibfnamefont {A.}~\bibnamefont {Vishwanath}},
  \ and\ \bibinfo {author} {\bibfnamefont {D.~A.}\ \bibnamefont {Huse}},\
  }\href {\doibase 10.1103/PhysRevX.8.031057} {\bibfield  {journal} {\bibinfo
  {journal} {Phys. Rev. X}\ }\textbf {\bibinfo {volume} {8}},\ \bibinfo {pages}
  {031057} (\bibinfo {year} {2018})}\BibitemShut {NoStop}%
\bibitem [{\citenamefont {Rakovszky}\ \emph {et~al.}(2018)\citenamefont
  {Rakovszky}, \citenamefont {Pollmann},\ and\ \citenamefont {von
  Keyserlingk}}]{rakovszky2018diffusive}%
  \BibitemOpen
  \bibfield  {author} {\bibinfo {author} {\bibfnamefont {T.}~\bibnamefont
  {Rakovszky}}, \bibinfo {author} {\bibfnamefont {F.}~\bibnamefont {Pollmann}},
  \ and\ \bibinfo {author} {\bibfnamefont {C.~W.}\ \bibnamefont {von
  Keyserlingk}},\ }\href {\doibase 10.1103/PhysRevX.8.031058} {\bibfield
  {journal} {\bibinfo  {journal} {Phys. Rev. X}\ }\textbf {\bibinfo {volume}
  {8}},\ \bibinfo {pages} {031058} (\bibinfo {year} {2018})}\BibitemShut
  {NoStop}%
\bibitem [{\citenamefont {Hunter-Jones}(2018)}]{hunter2018operator}%
  \BibitemOpen
  \bibfield  {author} {\bibinfo {author} {\bibfnamefont {N.}~\bibnamefont
  {Hunter-Jones}},\ }\href {https://arxiv.org/abs/1812.08219} {\bibfield
  {journal} {\bibinfo  {journal} {arXiv:1812.08219}\ } (\bibinfo {year}
  {2018})}\BibitemShut {NoStop}%
\bibitem [{\citenamefont {Hunter-Jones}(2019)}]{hunter2019unitary}%
  \BibitemOpen
  \bibfield  {author} {\bibinfo {author} {\bibfnamefont {N.}~\bibnamefont
  {Hunter-Jones}},\ }\href {https://arxiv.org/abs/1905.12053} {\bibfield
  {journal} {\bibinfo  {journal} {arXiv:1905.12053}\ } (\bibinfo {year}
  {2019})}\BibitemShut {NoStop}%
\bibitem [{\citenamefont {Zhou}\ and\ \citenamefont
  {Nahum}(2019)}]{zhou2019emergent}%
  \BibitemOpen
  \bibfield  {author} {\bibinfo {author} {\bibfnamefont {T.}~\bibnamefont
  {Zhou}}\ and\ \bibinfo {author} {\bibfnamefont {A.}~\bibnamefont {Nahum}},\
  }\href {\doibase 10.1103/PhysRevB.99.174205} {\bibfield  {journal} {\bibinfo
  {journal} {Phys. Rev. B}\ }\textbf {\bibinfo {volume} {99}},\ \bibinfo
  {pages} {174205} (\bibinfo {year} {2019})}\BibitemShut {NoStop}%
\bibitem [{\citenamefont {Bertini}\ and\ \citenamefont
  {Piroli}(2020)}]{bertini2020scrambling}%
  \BibitemOpen
  \bibfield  {author} {\bibinfo {author} {\bibfnamefont {B.}~\bibnamefont
  {Bertini}}\ and\ \bibinfo {author} {\bibfnamefont {L.}~\bibnamefont
  {Piroli}},\ }\href {\doibase 10.1103/PhysRevB.102.064305} {\bibfield
  {journal} {\bibinfo  {journal} {Phys. Rev. B}\ }\textbf {\bibinfo {volume}
  {102}},\ \bibinfo {pages} {064305} (\bibinfo {year} {2020})}\BibitemShut
  {NoStop}%
\bibitem [{\citenamefont {Zanardi}(2001)}]{zanardi2001entanglement}%
  \BibitemOpen
  \bibfield  {author} {\bibinfo {author} {\bibfnamefont {P.}~\bibnamefont
  {Zanardi}},\ }\href {\doibase 10.1103/PhysRevA.63.040304} {\bibfield
  {journal} {\bibinfo  {journal} {Phys. Rev. A}\ }\textbf {\bibinfo {volume}
  {63}},\ \bibinfo {pages} {040304} (\bibinfo {year} {2001})}\BibitemShut
  {NoStop}%
\bibitem [{\citenamefont {Prosen}\ and\ \citenamefont
  {Pi{\v{z}}orn}(2007)}]{prosen2007operator}%
  \BibitemOpen
  \bibfield  {author} {\bibinfo {author} {\bibfnamefont {T.}~\bibnamefont
  {Prosen}}\ and\ \bibinfo {author} {\bibfnamefont {I.}~\bibnamefont
  {Pi{\v{z}}orn}},\ }\href {\doibase 10.1103/PhysRevA.76.032316} {\bibfield
  {journal} {\bibinfo  {journal} {Phys. Rev. A}\ }\textbf {\bibinfo {volume}
  {76}},\ \bibinfo {pages} {032316} (\bibinfo {year} {2007})}\BibitemShut
  {NoStop}%
\bibitem [{\citenamefont {Dubail}(2017)}]{dubail2017entanglement}%
  \BibitemOpen
  \bibfield  {author} {\bibinfo {author} {\bibfnamefont {J.}~\bibnamefont
  {Dubail}},\ }\href {\doibase 10.1103/PhysRevA.76.032316} {\bibfield
  {journal} {\bibinfo  {journal} {J. Phys. A: Math. Theor.}\ }\textbf {\bibinfo
  {volume} {50}},\ \bibinfo {pages} {234001} (\bibinfo {year}
  {2017})}\BibitemShut {NoStop}%
\bibitem [{\citenamefont {Zhou}\ and\ \citenamefont
  {Luitz}(2017)}]{zhou2017operator}%
  \BibitemOpen
  \bibfield  {author} {\bibinfo {author} {\bibfnamefont {T.}~\bibnamefont
  {Zhou}}\ and\ \bibinfo {author} {\bibfnamefont {D.~J.}\ \bibnamefont
  {Luitz}},\ }\href {\doibase 10.1103/PhysRevB.95.094206} {\bibfield  {journal}
  {\bibinfo  {journal} {Phys. Rev. B}\ }\textbf {\bibinfo {volume} {95}},\
  \bibinfo {pages} {094206} (\bibinfo {year} {2017})}\BibitemShut {NoStop}%
\bibitem [{Note4()}]{Note4}%
  \BibitemOpen
  \bibinfo {note} {For RUC, it was proven that the two quantities coincide up
  to order $1/(d^8 \protect \qopname \relax o{ln}d)$~\cite
  {zhou2019emergent}.}\BibitemShut {Stop}%
\bibitem [{\citenamefont {Hosur}\ \emph {et~al.}(2016)\citenamefont {Hosur},
  \citenamefont {Qi}, \citenamefont {Roberts},\ and\ \citenamefont
  {Yoshida}}]{hosur2016chaos}%
  \BibitemOpen
  \bibfield  {author} {\bibinfo {author} {\bibfnamefont {P.}~\bibnamefont
  {Hosur}}, \bibinfo {author} {\bibfnamefont {X.-L.}\ \bibnamefont {Qi}},
  \bibinfo {author} {\bibfnamefont {D.~A.}\ \bibnamefont {Roberts}}, \ and\
  \bibinfo {author} {\bibfnamefont {B.}~\bibnamefont {Yoshida}},\ }\href
  {\doibase 10.1007/JHEP02(2016)004} {\bibfield  {journal} {\bibinfo  {journal}
  {JHEP}\ }\textbf {\bibinfo {volume} {2016}},\ \bibinfo {pages} {1} (\bibinfo
  {year} {2016})}\BibitemShut {NoStop}%
\bibitem [{\citenamefont {Li}\ \emph {et~al.}(2018)\citenamefont {Li},
  \citenamefont {Chen},\ and\ \citenamefont {Fisher}}]{yaodong2018quantum}%
  \BibitemOpen
  \bibfield  {author} {\bibinfo {author} {\bibfnamefont {Y.}~\bibnamefont
  {Li}}, \bibinfo {author} {\bibfnamefont {X.}~\bibnamefont {Chen}}, \ and\
  \bibinfo {author} {\bibfnamefont {M.~P.~A.}\ \bibnamefont {Fisher}},\ }\href
  {\doibase 10.1103/PhysRevB.98.205136} {\bibfield  {journal} {\bibinfo
  {journal} {Phys. Rev. B}\ }\textbf {\bibinfo {volume} {98}},\ \bibinfo
  {pages} {205136} (\bibinfo {year} {2018})}\BibitemShut {NoStop}%
\bibitem [{\citenamefont {Skinner}\ \emph {et~al.}(2019)\citenamefont
  {Skinner}, \citenamefont {Ruhman},\ and\ \citenamefont
  {Nahum}}]{skinner2019measurement}%
  \BibitemOpen
  \bibfield  {author} {\bibinfo {author} {\bibfnamefont {B.}~\bibnamefont
  {Skinner}}, \bibinfo {author} {\bibfnamefont {J.}~\bibnamefont {Ruhman}}, \
  and\ \bibinfo {author} {\bibfnamefont {A.}~\bibnamefont {Nahum}},\ }\href
  {\doibase 10.1103/PhysRevX.9.031009} {\bibfield  {journal} {\bibinfo
  {journal} {Phys. Rev. X}\ }\textbf {\bibinfo {volume} {9}},\ \bibinfo {pages}
  {031009} (\bibinfo {year} {2019})}\BibitemShut {NoStop}%
\bibitem [{\citenamefont {Chan}\ \emph
  {et~al.}(2019{\natexlab{b}})\citenamefont {Chan}, \citenamefont
  {Nandkishore}, \citenamefont {Pretko},\ and\ \citenamefont
  {Smith}}]{chan2019unitary}%
  \BibitemOpen
  \bibfield  {author} {\bibinfo {author} {\bibfnamefont {A.}~\bibnamefont
  {Chan}}, \bibinfo {author} {\bibfnamefont {R.~M.}\ \bibnamefont
  {Nandkishore}}, \bibinfo {author} {\bibfnamefont {M.}~\bibnamefont {Pretko}},
  \ and\ \bibinfo {author} {\bibfnamefont {G.}~\bibnamefont {Smith}},\ }\href
  {\doibase 10.1103/PhysRevB.99.224307} {\bibfield  {journal} {\bibinfo
  {journal} {Phys. Rev. B}\ }\textbf {\bibinfo {volume} {99}},\ \bibinfo
  {pages} {224307} (\bibinfo {year} {2019}{\natexlab{b}})}\BibitemShut
  {NoStop}%
\bibitem [{\citenamefont {Fan}\ \emph {et~al.}(2020)\citenamefont {Fan},
  \citenamefont {Vijay}, \citenamefont {Vishwanath},\ and\ \citenamefont
  {You}}]{fan2020self}%
  \BibitemOpen
  \bibfield  {author} {\bibinfo {author} {\bibfnamefont {R.}~\bibnamefont
  {Fan}}, \bibinfo {author} {\bibfnamefont {S.}~\bibnamefont {Vijay}}, \bibinfo
  {author} {\bibfnamefont {A.}~\bibnamefont {Vishwanath}}, \ and\ \bibinfo
  {author} {\bibfnamefont {Y.-Z.}\ \bibnamefont {You}},\ }\href
  {https://arxiv.org/abs/2002.12385} {\bibfield  {journal} {\bibinfo  {journal}
  {arXiv:2002.12385}\ } (\bibinfo {year} {2020})}\BibitemShut {NoStop}%
\bibitem [{\citenamefont {Choi}\ \emph {et~al.}(2020)\citenamefont {Choi},
  \citenamefont {Bao}, \citenamefont {Qi},\ and\ \citenamefont
  {Altman}}]{choi2020quantum}%
  \BibitemOpen
  \bibfield  {author} {\bibinfo {author} {\bibfnamefont {S.}~\bibnamefont
  {Choi}}, \bibinfo {author} {\bibfnamefont {Y.}~\bibnamefont {Bao}}, \bibinfo
  {author} {\bibfnamefont {X.-L.}\ \bibnamefont {Qi}}, \ and\ \bibinfo {author}
  {\bibfnamefont {E.}~\bibnamefont {Altman}},\ }\href {\doibase
  10.1103/PhysRevLett.125.030505} {\bibfield  {journal} {\bibinfo  {journal}
  {Phys. Rev. Lett.}\ }\textbf {\bibinfo {volume} {125}},\ \bibinfo {pages}
  {030505} (\bibinfo {year} {2020})}\BibitemShut {NoStop}%
\bibitem [{\citenamefont {Gullans}\ and\ \citenamefont
  {Huse}(2020)}]{gullans2020dynamical}%
  \BibitemOpen
  \bibfield  {author} {\bibinfo {author} {\bibfnamefont {M.~J.}\ \bibnamefont
  {Gullans}}\ and\ \bibinfo {author} {\bibfnamefont {D.~A.}\ \bibnamefont
  {Huse}},\ }\href {\doibase 10.1103/PhysRevX.10.041020} {\bibfield  {journal}
  {\bibinfo  {journal} {Phys. Rev. X}\ }\textbf {\bibinfo {volume} {10}},\
  \bibinfo {pages} {041020} (\bibinfo {year} {2020})}\BibitemShut {NoStop}%
\bibitem [{\citenamefont {Ippoliti}\ \emph {et~al.}(2021)\citenamefont
  {Ippoliti}, \citenamefont {Gullans}, \citenamefont {Gopalakrishnan},
  \citenamefont {Huse},\ and\ \citenamefont
  {Khemani}}]{ippoliti2021entanglement}%
  \BibitemOpen
  \bibfield  {author} {\bibinfo {author} {\bibfnamefont {M.}~\bibnamefont
  {Ippoliti}}, \bibinfo {author} {\bibfnamefont {M.~J.}\ \bibnamefont
  {Gullans}}, \bibinfo {author} {\bibfnamefont {S.}~\bibnamefont
  {Gopalakrishnan}}, \bibinfo {author} {\bibfnamefont {D.~A.}\ \bibnamefont
  {Huse}}, \ and\ \bibinfo {author} {\bibfnamefont {V.}~\bibnamefont
  {Khemani}},\ }\href {\doibase 10.1103/PhysRevX.11.011030} {\bibfield
  {journal} {\bibinfo  {journal} {Phys. Rev. X}\ }\textbf {\bibinfo {volume}
  {11}},\ \bibinfo {pages} {011030} (\bibinfo {year} {2021})}\BibitemShut
  {NoStop}%
\bibitem [{\citenamefont {Jian}\ \emph {et~al.}(2020)\citenamefont {Jian},
  \citenamefont {You}, \citenamefont {Vasseur},\ and\ \citenamefont
  {Ludwig}}]{jian2020measurement}%
  \BibitemOpen
  \bibfield  {author} {\bibinfo {author} {\bibfnamefont {C.-M.}\ \bibnamefont
  {Jian}}, \bibinfo {author} {\bibfnamefont {Y.-Z.}\ \bibnamefont {You}},
  \bibinfo {author} {\bibfnamefont {R.}~\bibnamefont {Vasseur}}, \ and\
  \bibinfo {author} {\bibfnamefont {A.~W.~W.}\ \bibnamefont {Ludwig}},\ }\href
  {\doibase 10.1103/PhysRevB.101.104302} {\bibfield  {journal} {\bibinfo
  {journal} {Phys. Rev. B}\ }\textbf {\bibinfo {volume} {101}},\ \bibinfo
  {pages} {104302} (\bibinfo {year} {2020})}\BibitemShut {NoStop}%
\bibitem [{\citenamefont {Haah}\ \emph {et~al.}(2018)\citenamefont {Haah},
  \citenamefont {Fidkowski},\ and\ \citenamefont
  {Hastings}}]{haah2018nontrivial}%
  \BibitemOpen
  \bibfield  {author} {\bibinfo {author} {\bibfnamefont {J.}~\bibnamefont
  {Haah}}, \bibinfo {author} {\bibfnamefont {L.}~\bibnamefont {Fidkowski}}, \
  and\ \bibinfo {author} {\bibfnamefont {M.~B.}\ \bibnamefont {Hastings}},\
  }\href {https://arxiv.org/abs/1812.01625} {\bibfield  {journal} {\bibinfo
  {journal} {arXiv:1812.01625}\ } (\bibinfo {year} {2018})}\BibitemShut
  {NoStop}%
\bibitem [{\citenamefont {Haah}(2019)}]{haah2019clifford}%
  \BibitemOpen
  \bibfield  {author} {\bibinfo {author} {\bibfnamefont {J.}~\bibnamefont
  {Haah}},\ }\href {https://arxiv.org/abs/1907.02075} {\bibfield  {journal}
  {\bibinfo  {journal} {arXiv:1907.02075}\ } (\bibinfo {year}
  {2019})}\BibitemShut {NoStop}%
\bibitem [{\citenamefont {Freedman}\ \emph {et~al.}(2019)\citenamefont
  {Freedman}, \citenamefont {Haah},\ and\ \citenamefont
  {Hastings}}]{freedman2019group}%
  \BibitemOpen
  \bibfield  {author} {\bibinfo {author} {\bibfnamefont {M.}~\bibnamefont
  {Freedman}}, \bibinfo {author} {\bibfnamefont {J.}~\bibnamefont {Haah}}, \
  and\ \bibinfo {author} {\bibfnamefont {M.~B.}\ \bibnamefont {Hastings}},\
  }\href {https://arxiv.org/abs/1910.07998} {\bibfield  {journal} {\bibinfo
  {journal} {arXiv:1910.07998}\ } (\bibinfo {year} {2019})}\BibitemShut
  {NoStop}%
\bibitem [{\citenamefont {Freedman}\ and\ \citenamefont
  {Hastings}(2020)}]{freedman2020classification}%
  \BibitemOpen
  \bibfield  {author} {\bibinfo {author} {\bibfnamefont {M.}~\bibnamefont
  {Freedman}}\ and\ \bibinfo {author} {\bibfnamefont {M.~B.}\ \bibnamefont
  {Hastings}},\ }\href {\doibase 10.1007/s00220-020-03735-y} {\bibfield
  {journal} {\bibinfo  {journal} {Comm. Math. Phys.}\ }\textbf {\bibinfo
  {volume} {376}},\ \bibinfo {pages} {1171} (\bibinfo {year}
  {2020})}\BibitemShut {NoStop}%
\bibitem [{\citenamefont {Paris}(2013{\natexlab{a}})}]{paris2013asymptotics}%
  \BibitemOpen
  \bibfield  {author} {\bibinfo {author} {\bibfnamefont {R.~B.}\ \bibnamefont
  {Paris}},\ }\href {\doibase 10.7153/jca-02-15} {\bibfield  {journal}
  {\bibinfo  {journal} {J. Class. Analysis}\ }\textbf {\bibinfo {volume} {2}},\
  \bibinfo {pages} {183} (\bibinfo {year} {2013}{\natexlab{a}})}\BibitemShut
  {NoStop}%
\bibitem [{\citenamefont {Paris}(2013{\natexlab{b}})}]{paris2013asymptoticsII}%
  \BibitemOpen
  \bibfield  {author} {\bibinfo {author} {\bibfnamefont {R.~B.}\ \bibnamefont
  {Paris}},\ }\href
  {https://rke.abertay.ac.uk/en/publications/asymptotics-of-the-gauss-hypergeometric-function-with-large-param-2}
  {\bibfield  {journal} {\bibinfo  {journal} {J. Class. Analysis}\ }\textbf
  {\bibinfo {volume} {3}},\ \bibinfo {pages} {1} (\bibinfo {year}
  {2013}{\natexlab{b}})}\BibitemShut {NoStop}%
\bibitem [{\citenamefont {S{\"u}nderhauf}\ \emph {et~al.}(2019)\citenamefont
  {S{\"u}nderhauf}, \citenamefont {Piroli}, \citenamefont {Qi}, \citenamefont
  {Schuch},\ and\ \citenamefont {Cirac}}]{sunderhauf2019quantum}%
  \BibitemOpen
  \bibfield  {author} {\bibinfo {author} {\bibfnamefont {C.}~\bibnamefont
  {S{\"u}nderhauf}}, \bibinfo {author} {\bibfnamefont {L.}~\bibnamefont
  {Piroli}}, \bibinfo {author} {\bibfnamefont {X.-L.}\ \bibnamefont {Qi}},
  \bibinfo {author} {\bibfnamefont {N.}~\bibnamefont {Schuch}}, \ and\ \bibinfo
  {author} {\bibfnamefont {J.~I.}\ \bibnamefont {Cirac}},\ }\href {\doibase
  10.1007/JHEP11(2019)038} {\bibfield  {journal} {\bibinfo  {journal} {JHEP}\
  }\textbf {\bibinfo {volume} {2019}},\ \bibinfo {pages} {38} (\bibinfo {year}
  {2019})}\BibitemShut {NoStop}%
\bibitem [{\citenamefont {Schnaack}\ \emph {et~al.}(2019)\citenamefont
  {Schnaack}, \citenamefont {B\"olter}, \citenamefont {Paeckel}, \citenamefont
  {Manmana}, \citenamefont {Kehrein},\ and\ \citenamefont
  {Schmitt}}]{schnaack2019tripartite}%
  \BibitemOpen
  \bibfield  {author} {\bibinfo {author} {\bibfnamefont {O.}~\bibnamefont
  {Schnaack}}, \bibinfo {author} {\bibfnamefont {N.}~\bibnamefont {B\"olter}},
  \bibinfo {author} {\bibfnamefont {S.}~\bibnamefont {Paeckel}}, \bibinfo
  {author} {\bibfnamefont {S.~R.}\ \bibnamefont {Manmana}}, \bibinfo {author}
  {\bibfnamefont {S.}~\bibnamefont {Kehrein}}, \ and\ \bibinfo {author}
  {\bibfnamefont {M.}~\bibnamefont {Schmitt}},\ }\href {\doibase
  10.1103/PhysRevB.100.224302} {\bibfield  {journal} {\bibinfo  {journal}
  {Phys. Rev. B}\ }\textbf {\bibinfo {volume} {100}},\ \bibinfo {pages}
  {224302} (\bibinfo {year} {2019})}\BibitemShut {NoStop}%
\bibitem [{\citenamefont {Kudler-Flam}\ \emph {et~al.}(2021)\citenamefont
  {Kudler-Flam}, \citenamefont {Sohal},\ and\ \citenamefont
  {Nie}}]{kudler2021information}%
  \BibitemOpen
  \bibfield  {author} {\bibinfo {author} {\bibfnamefont {J.}~\bibnamefont
  {Kudler-Flam}}, \bibinfo {author} {\bibfnamefont {R.}~\bibnamefont {Sohal}},
  \ and\ \bibinfo {author} {\bibfnamefont {L.}~\bibnamefont {Nie}},\ }\href
  {https://arxiv.org/abs/2107.04043} {\bibfield  {journal} {\bibinfo  {journal}
  {arXiv:2107.04043}\ } (\bibinfo {year} {2021})}\BibitemShut {NoStop}%
\end{thebibliography}%

\clearpage
\onecolumngrid
\begin{center}
	\textbf{\large Supplemental Materials}
\end{center}
\setcounter{equation}{0}
\setcounter{figure}{0}
\setcounter{table}{0}
\makeatletter
\renewcommand{\theequation}{S\arabic{equation}}
\renewcommand{\thefigure}{S\arabic{figure}}
\renewcommand{\bibnumfmt}[1]{[S#1]}

We provide the detailed derivations of various dynamical quantities for the exactly solvable random QCA in the main text. We also discuss a different simplified random QCA and the parent Floquet systems of both.

\section{Models of random QCA}

In the main text, we have considered a natural model of random QCA. It is defined on a chain with periodic boundary conditions, associated with the Hilbert  space $(\mathbb{C}^{d})^{\otimes 2L}$ ($2L$: system size). The unitary dynamics is discrete, and dictated by the evolution operator
\begin{equation}
{\cal U}(t)=  \overleftarrow{\prod^{t-1}_{s=0}}U_s= \overleftarrow{\prod^{t-1}_{s=0}}\left( \bigotimes^{L-1}_{j=0}v^{[2j-1,2j]}_{s+\frac{1}{2}}\right) \left( \bigotimes^{L-1}_{j=0}u^{[2j,2j+1]}_s\right)\,.
\label{eq:stdQCA}
\end{equation} 
Here $u:\mathbb{C}^d\otimes \mathbb{C}^d \to \mathbb{C}^p\otimes \mathbb{C}^q$ and $v: \mathbb{C}^q\otimes \mathbb{C}^p\to\mathbb{C}^d\otimes \mathbb{C}^d$ are unitary operators, where $p$ and $q$ integers such that $pq=d^2$. In our model, $u$ and $v$ are drawn randomly from the Haar distribution and independently for each point in space and time. If $p\neq d$, then the dynamics is anomalous, and is characterized by a nonzero index~\cite{gross2012index}. Here, we omit its mathematical definition and simply recall that for the time-step evolution operator $U_s$ in~\eqref{eq:stdQCA} it is easily computed and reads
\begin{equation}
	\ind = \frac{1}{2} \ln \frac{q}{p}.
	\label{eq:index_model_A}
\end{equation}

\begin{figure*}
	\begin{center}
		\includegraphics[scale=0.28]{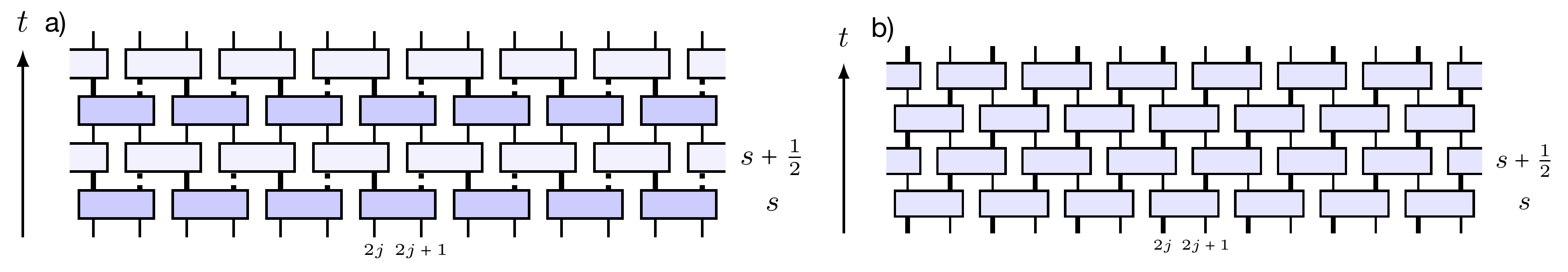}
	\end{center}
	\caption{Models of random QCA. a) Model A: The unitary evolution operator is given in Eq.~\eqref{eq:stdQCA}. For $p\neq d$ the dynamics is anomalous, with index~\eqref{eq:index_model_A}. b) Model B:  The dynamics is dictated by the single-step evolution operator $U_{s}=(\otimes^{L-1}_{j=0} u^{[2j-1,2j]}_{s+\frac{1}{2}}) (\otimes^{L-1}_{j=0} u^{[2j,2j+1]}_s)$ where $u^{[a,a+1]}_s: \mathbb{C}^p\otimes\mathbb{C}^q\to \mathbb{C}^q\otimes\mathbb{C}^p$ are unitaries drawn randomly from the Haar distribution, and independently for each point in space and time. For $p\neq q$  the dynamics is anomalous, with index~\eqref{eq:index_model_B}.}
	\label{fig:models_qca}
\end{figure*}

In fact, it is possible to construct other simple models of anomalous dynamics. Indeed it is known that any discrete unitary evolution with a given index may be obtained by combining layers of quantum circuits and shifts~\cite{gross2012index}. When $d=pq$, where $p$ and $q$ are positive integers, the local Hilbert space factorizes as $\mathbb{C}^d=\mathbb{C}^p\otimes \mathbb{C}^q$, and one may define two shift operators $T_{p}$ and $T_{q}$ translating to the right only one of the two types of local subspaces. Then, the operator $\mathbb{T}_{p,q}=T^{-1}_qT_p$ has nonzero index ${\rm ind}=\ln(p/q)$~\cite{gross2012index}. The dynamics generated by $\mathbb{T}_{p,q}
$ is clearly not generic. However, in order to obtain a simple chaotic model we may alternate this shift with a direct product of on-site Haar random unitaries $u$, yielding
\begin{equation}\label{eq:u_simplified}
	{\cal U}(t)= \overleftarrow{\prod^{t-1}_{s=0}}\left( \mathbb{T}_{p,q} \bigotimes^{L-1}_{j=0}u^{[j]}_s\right)\,,
\end{equation}
where $u$ are drawn randomly from the Haar distribution and independently for each point in space and time. ${\cal U}(t)$ then admits the graphical representation
\begin{equation}
	\begin{tikzpicture}
		\foreach \x in {0,...,6}
		\foreach \y in {0,...,1}
		{
			\draw[thick,fill=blue!10!white] (\x-0.4,\y-0.1) rectangle (\x+0.4,\y-0.4); 
			\draw[ultra thick] (\x-0.25,\y-0.4) -- (\x-0.25,\y-0.6) (\x-0.25,\y-0.1) -- (\x-0.25,\y+0.1);
			\draw[thick] (\x+0.25,\y-0.4) -- (\x+0.25,\y-0.6) (\x+0.25,\y-0.1) -- (\x+0.25,\y+0.1);
		}
		\foreach \x in {0,...,5}
		\foreach \y in {0,...,1}
		{
			\draw[ultra thick] (\x-0.25,\y+0.1) .. controls (\x-0.25,\y+0.3) and (\x+0.75,\y+0.2) .. (\x+0.75,\y+0.4);
			\draw[thick] (\x+0.25,\y+0.4) .. controls (\x+0.25,\y+0.2) and (\x+1.25,\y+0.3) ..  (\x+1.25,\y+0.1);
		}
		\foreach \y in {0,...,1}
		{
			\draw[ultra thick] (-0.5,\y+0.25) .. controls (-0.25,\y+0.3) .. (-0.25,\y+0.4) (5.75,\y+0.1) .. controls (5.75,\y+0.2) .. (6.5,\y+0.25);
			\draw[thick] (-0.5,\y+0.25) .. controls (0.25,\y+0.2) .. (0.25,\y+0.1) (6.25,\y+0.4) .. controls (6.25,\y+0.3) .. (6.5,\y+0.25);
		}
		\foreach \x in {0,...,6}
		{
			\draw[ultra thick]  (\x-0.25,1.4) -- (\x-0.25,1.6);
			\draw[thick] (\x+0.25,1.4) -- (\x+0.25,1.6);
		}
		\begin{scope}[>=latex] 
			\draw[->,thick] (-1,-0.6) -- (-1,1.4);
		\end{scope}
		\Text[x=-1,y=1.7]{$t$}
		\Text[x=3,y=-0.75,fontsize=\small]{$j$}
		\Text[x=7,y=-0.25,fontsize=\small]{$s$}
		\Text[x=7,y=0.25,fontsize=\small]{$s+\frac{1}{2}$}
	\end{tikzpicture}
	\label{uT}
\end{equation}
where the dimension of thick (thin) legs is $p$ $(q)$. We see that blocks with even $s+j$ are fully decoupled from those with odd $s+j$. Also, due to the properties of the Haar measure, we have
\begin{equation}
	\begin{tikzpicture}
		\draw[thick,fill=blue!10!white] (-0.4,-0.2) rectangle (0.4,0.1);
		\draw[ultra thick] (-0.25,-0.4) -- (-0.25,-0.2) (-0.25,0.1) -- (-0.25,0.2) (-0.25,0.2) .. controls (-0.25,0.3) and (0.25,0.3) .. (0.25,0.4);
		\draw[thick] (0.25,-0.4) -- (0.25,-0.2) (0.25,0.1) -- (0.25,0.2) (0.25,0.2) .. controls (0.25,0.3) and (-0.25,0.3) .. (-0.25,0.4);
		\Text[x=0.8,y=0]{$=$}
		\draw[thick,fill=blue!10!white] (1.2,-0.2) rectangle (2,0.1);
		\draw[ultra thick] (1.35,-0.4) -- (1.35,-0.2) (1.85,0.1) -- (1.85,0.3);
		\draw[thick] (1.85,-0.4) -- (1.85,-0.2) (1.35,0.1) -- (1.35,0.3);
	\end{tikzpicture}
\end{equation}
at the level of random ensembles. Therefore, this construction leads to two identical copies of a simplified model of random QCA, reported in Fig.~\ref{fig:models_qca}b), with index
\begin{equation}
	\ind =\ln \frac{p}{q}\,.
	\label{eq:index_model_B}
\end{equation}
We will call this \emph{Model B}, in order to distinguish it from the one defined by~\eqref{eq:stdQCA} [and depicted in Fig.~\ref{fig:models_qca}a)], which we will call  \emph{Model A}. 

In order to verify the generality of the proposed EMT, we have performed analytic computations in both models. In fact, the calculations
are very similar in the two cases. Therefore, we will detail them only for Model A, for which they are slightly more complicated, while for Model B we will merely report the final results.

\section{Details on the parent Floquet systems}

Here we briefly discuss  the $2$D parent Floquet dynamics for the random QCA studied in this work. They can be easily defined for both models A and B. 

\begin{figure*}
	\begin{center}
		\includegraphics[width=16cm, clip]{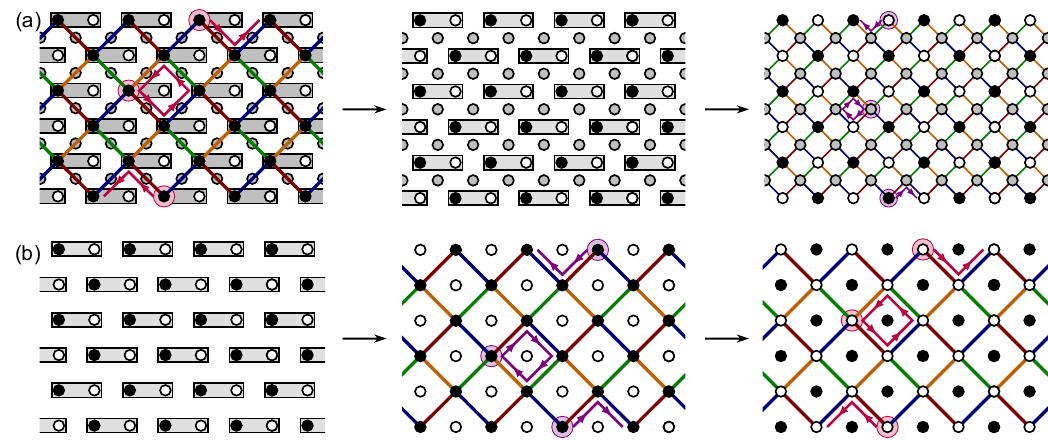}
	\end{center}
	\caption{Parent Floquet systems for (a) Model A and (b) Model B. (a) Compared to Ref.~\cite{gong2020classification}, there are two main differences: (i) In the first step, the bipartite unitaries that map (by conjugation) physical degrees of freedom to virtual ones are randomized; (ii) there is an additional middle step of applying randomized two-site gates on the physical level. (b) After an on-site randomization, sublattices $\Lambda_p$ (black circles) and $\Lambda_q$ (white circles) undergo clockwise and counterclockwise four-step swap operations (following the order \textcolor{red!50!black}{red}$\to$\textcolor{blue!50!black}{blue}$\to$\textcolor{green!50!black}{green}$\to$\textcolor{orange!75!black}{orange}; same in (a)), respectively. The operations on the two sublattices can be performed simultaneously.}
	\label{fig:FPH}
\end{figure*}

We begin by analyzing Model B, for which the construction can be straightforwardly obtained from the results presented in Ref.~\cite{po2016chiral}. We consider a rectangular system, in which we place a lattice $\Lambda$, rotated $45$ degrees with respect to a horizontal axis. 
The local  Hilbert space is $\mathbb{C}^{d}=\mathbb{C}^{p}\otimes \mathbb{C}^{q}$, each factor forming a sublattice denoted by $\Lambda_p$ and $\Lambda_q$, respectively. As shown in~\cite{po2016chiral}, for $\Lambda_p$  ($\Lambda_q$) it is easy to construct a quantum circuit of depth $4$ made of swaps which implements a shift of a full period on each plaquette 
[cf. Fig.~\ref{fig:FPH}(b)]. In turn, such quantum circuit can be clearly defined in terms of finite-time evolution of a local Floquet Hamiltonian. The ensuing dynamics is trivial in the bulk, while it implements a non-trivial shift at the boundaries. Choosing opposite directions for the shifts in $\Lambda_p$ and $\Lambda_q$, the (bottom) boundary Floquet operator reads $U_F= \mathbb{T}_{p,q}=T_q^{-1}T_p$,  where $T_p$ and $T_{q}$ are translations to the right acting on the two types of local subspaces. We see that in order to obtain~\eqref{eq:u_simplified}, it is then enough to apply a unitary $V=\bigotimes_{\boldsymbol{r}}u_{\boldsymbol{r}}$, where $u_{\boldsymbol{r}}$ is a random one-site operator acting on the local space $\mathbb{C}^d$ at position $\boldsymbol{r}$. Clearly, $V$ may be obtained as finite-time evolution of a random local Hamiltonian. Furthermore, $V$ does not spoil localization in the bulk. This completes the construction of the parent Floquet  system for Model B.

Similarly, we can construct a parent Floquet evolution for Model A. This is done following a construction presented in~\cite{gong2020classification}, which we briefly sketch here for completeness. We again consider a square lattice with local Hilbert space $\mathbb{C}^d$. As shown in the left panel in Fig.~\ref{fig:FPH}, we perform the four-step-swap operation on the virtual level, which is pulled back to the physical level by a conjugation by disjoint random unitaries $u: \mathbb{C}^d\otimes\mathbb{C}^d\to \mathbb{C}^p\otimes\mathbb{C}^q$. According to Ref.~\cite{gong2020classification}, if this is followed by a four-step-swap operation on the physical level, as is shown in the right panel, the edge dynamics will be a QCA in the Margolous form with $v^{[2j+1,2j+2]}=u^{[2j,2j+1]\dag}\mathbb{S}_{p,q}$, where $\mathbb{S}_{p,q}$ swaps $\mathbb{C}^q$ and $\mathbb{C}^p$. To make $v$'s completely random, i.e., independent of $u$'s, we only have to apply a layer of random two-site gates on the physical level, as shown in the middle panel in Fig.~\ref{fig:FPH}(a). Again, these disjoint local unitaries do not spoil the many-body localizability in the bulk. This completes the construction of the parent Floquet  system for Model A.

\section{Mapping to an Ising partition function: the OTOCs}

Here we provide details on the calculations of the OTOC
\begin{equation}\label{eq:OTOC_SM}
	\mathcal{C}(x, t) \equiv \frac{1}{2} \operatorname{tr}( \left[O_{0}, O^\prime_{2x}(t)\right]^{\dagger}\left[O_{0}, O^\prime_{2x}(t)\right])\,,
\end{equation}
where $O_{0}$, $O^\prime_{2x}$ are arbitrary traceless operator satisfying $O^2=O^{\prime2}=\openone$, and $O^\prime_{2x}(t)=\mathcal{U}(t)^\dag O^\prime_{2x}\mathcal{U}(t)$. We used  the same approach developed in the case of RUC~\cite{nahum2017quantum,nahum2018operator,vonKeyserlingk2018operator}, which is based on a mapping onto the partition function of  a classical Ising-like model in $2$D. In fact, our calculations follow closely those presented in these works. For this reason, we simply sketch the main steps and work out the parts of the analysis which are different from the case of RUC. We will focus on Model A, and only present the final results for Model B at the end. 

The first step is to represent the quantities of interest in terms of the ``replica operator'' ${\cal U}^{(N)}(t)=[{\cal U}(t)\otimes {\cal U}^\ast(t)]^{\otimes N}$, where ${\cal U}^\ast(t)$ denotes complex conjugation with respect to the computation basis. Clearly, ${\cal U}^{(N)}(t)$ has the same geometrical structure of ${\cal U}(t)$, but in a ``replica space'' where the local Hilbert-space dimension is $d^{2N}$. For instance, in the case ${\cal U}(t)$ is a RUC with brickwork structure, so is ${\cal U}^{(N)}(t)$, with local two-site gates $u^{(N)}=(u\otimes u^\ast)^{\otimes N}$. This representation is very convenient because the unitaries $u^{(N)}$ are distributed independently from one another, and averages over disorder then factorize~\cite{nahum2017quantum,nahum2018operator,vonKeyserlingk2018operator}. After averaging, one is left with a non-unitary $1$D evolution, which in turn can be mapped onto a classical statistical-mechanics model of Ising spins in $2$D. The quantity under investigation determines the boundary conditions of this partition function. The latter may finally be evaluated exactly by solving a problem of domain-wall counting. We refer to Ref.~\cite{nahum2018operator} for the details. 

\subsection{Model A}
For the random QCA studied here, the very same approach can be applied. As the main difference, the QCA display staggered dynamics, which is reflected in the fact that the weights in the $2$D Ising model are different for even and odd rows. Although this makes the enumeration of domain walls more involved, the procedure is straightforward, and therefore we omit it. As a final result we obtain the following exact expression, valid at finite distances and times
\begin{align}\label{eq:exact_OTOC}
	{\cal C}(x,t)=\frac{d^4}{d^4-1}g_{L}\left(2t-1, t-x\right) g_{R}\left(2t-1, t+x\right) -\frac{1}{d^4-1}g_{L}\left(2t-1, t-x-1\right) g_{R}\left(2t-1, t+x-1\right) \,,
\end{align}
where
\begin{align}
	g_{L}(n,a)&=\sum_{u=0}^a f^{n}_{L}(u)d^{-2u} \left(\frac{d^2}{d^2+1}\right)^{n} \left[\frac{q(p^2-1)}{p(d^2 -1)}\right]^{(n+1)/2}
	\,,\label{eq:g_L}\\
	g_{R}(n,a)&=\sum_{u=0}^a f^{n}_{R}(u)d^{-2u} \left(\frac{d^2}{d^2+1}\right)^{n} \left[\frac{p(q^2-1)}{q(d^2 -1)}\right]^{(n+1)/2}
	\,,\label{eq:g_R}
\end{align}
and
\begin{align}
	f^{t-1}_{L}(x)&=\sum_{a=0}^x\binom{t/2}{a}\binom{t/2-1}{x-a}\left[\frac{p(q^2-1)}{q(p^2-1)}\right]^{a}\,,\\
	f^{t-1}_{R}(x)&=\sum_{a=0}^x\binom{t/2}{a}\binom{t/2-1}{x-a}\left[\frac{q(p^2-1)}{p(q^2-1)}\right]^{a}\,.
\end{align}
As a first check, we see that for $p=q=d$ we recover the result for the OTOC in RUC derived in Ref.~\cite{nahum2018operator}, after replacing $(x,t)=(\frac{x}{2}, \frac{t}{2})$: this can be seen using Vandermonde's identity, which gives $f^{t-1}_L(x)=f^{t-1}_R(x)=\binom{t-1}{x}$.

Let us now consider the OTOC in the limit of large spacetime distances. First, neglecting subleading terms, we may rewrite~\eqref{eq:exact_OTOC} as
\begin{align}\label{eq:exact_OTOC_large_t}
	{\cal C}(x,t)\simeq g_{L}\left(2t-1, t-x\right) g_{R}\left(2t-1, t+x\right) \,.
\end{align}
The analysis of~\eqref{eq:exact_OTOC_large_t} for large $x$ and $t$ is significantly more complicated than the case of RUC. This is because for $p\neq q$ the functions $f^{t-1}_L(x)$, $f^{t-1}_R(x)$ can not be expressed as simple binomial coefficients. By inspection, we see that the hydrodynamic limit of ${\cal C}(x,t)$ is determined by the terms in the sums~\eqref{eq:g_L}, \eqref{eq:g_R} where $u$ and $n$ are of the same order. Hence, we need an asymptotic expansion for $f^{2t-1}_{L/R}(u)$ when $u=\alpha t$ and $t\to\infty$. A systematic analysis of this limit may be obtained by expressing $f^{n}_{L/R}(x)$ in terms of the Gaussian hypergeometric function and exploiting known asymptotic expansion formulas~\cite{paris2013asymptotics,paris2013asymptoticsII}. Since we are only interested in the leading behavior in $t$, we follow a simpler and more direct approach. We begin by considering the function
\begin{equation}
	f(t,x,r)= \sum^x_{a=0}\binom{t}{a} \binom{t}{x-a}r^a\qquad  t,x\in\mathbb{N}\quad r\in\mathbb{R}^+\,,
\label{ftxr}
\end{equation}
which can be rewritten as
\begin{equation}
	f(t,x,r)=\oint_{C_z} \frac{dz}{2\pi i} (1+rz)^t (1+z)^t z^{-(1+x)},
	\label{fint}
\end{equation}
where $C_z$ can be an arbitrary closed loop encircling the origin. Next, we parametrize $z=\rho e^{ik}$, with $\rho,k\in \mathbb{R}$, and setting $x=\alpha t$, we obtain
\begin{equation}
	f(t,\alpha t,r)= \rho^{-x}\int_{0}^{2\pi}
	\frac{dk}{2\pi} \left[e^{-i \alpha k} (1+r\rho e^{ik})(1+\rho e^{ik})\right]^t\,.
	\label{fint2}
\end{equation}
For large $t$, we wish to evaluate this integral via the saddle-point method, which yields the saddle-point equation
\begin{equation}
	\rho^2r(2-\alpha)e^{2ik} + \rho(r+1)(1-\alpha)e^{ik} -\alpha=0,
\end{equation}
namely
\begin{equation}\label{eq:saddle_point_sol}
e^{ik}=\frac{-(1+r)(1-\alpha)\pm\sqrt{(1-r)^2(1-\alpha)^2+4r}}{2r\rho(2-\alpha)}\,.
\end{equation}
For $r$, $\alpha$ fixed, there is only one $\rho^\ast>0$ for which the solution to~\eqref{eq:saddle_point_sol} satisfies $k\in[0,2\pi]$. Therefore, we may choose $\rho=\rho^\ast$ and apply the saddle-point method. By examining the limit of $r=1$, we know that the dominant contribution should arise choosing the ``$+$" sign in~\eqref{eq:saddle_point_sol}. Substituting the corresponding root into Eq.~\eqref{fint2} yields
\begin{equation}
	\lim_{t\to\infty} \frac{ \ln f(t,\alpha t,r)}{t} = (1-\alpha) \ln\frac{-(1+r)(1-\alpha)+\sqrt{(1-r)^2(1-\alpha)^2+4r}}{2r(2-\alpha)} 
	+ \ln \frac{1 + r + \sqrt{(1-r)^2(1-\alpha)^2 + 4r}}{\alpha(2-\alpha)}\,,
	\label{fasym}
\end{equation}
which finally determines the leading exponential behavior of the function~\eqref{ftxr}. Next, to the  leading order in $t$, it is immediate to see that $f_{L/R}^{2t-1}(\alpha t)\simeq f(t,\alpha t,r_p^{\pm 1})$, where
\be\label{eq:rp}
r_{p}=\frac{p\left(q^{2}-1\right)}{q\left(p^{2}-1\right)}\,.
\ee
 Therefore, we may exploit the asymptotic expansion~\eqref{fasym} and obtain the final result
\be
f^{2t-1}_L(\alpha t)\simeq A(t,\alpha) \exp[t\mathcal{H}(\alpha, r_p)]\,,\qquad  f^{2t-1}_R(\alpha t)\simeq A(t,\alpha) \exp[t \mathcal{H}(\alpha, r^{-1}_p)]\,,
\ee
where
$A(t,\alpha)$ is a function growing sub-exponentially in $t$, and
\be
\mathcal{H}(\alpha, r)=(1-\alpha) \ln \frac{-(1+r)(1-\alpha)+\sqrt{(1-r)^{2}(1-\alpha)^{2}+4 r}}{2 r(2-\alpha)}+\ln \frac{1+r+\sqrt{(1-r)^{2}(1-\alpha)^{2}+4 r}}{\alpha(2-\alpha)}\,.
\ee

\begin{figure}
	\includegraphics[scale=0.46]{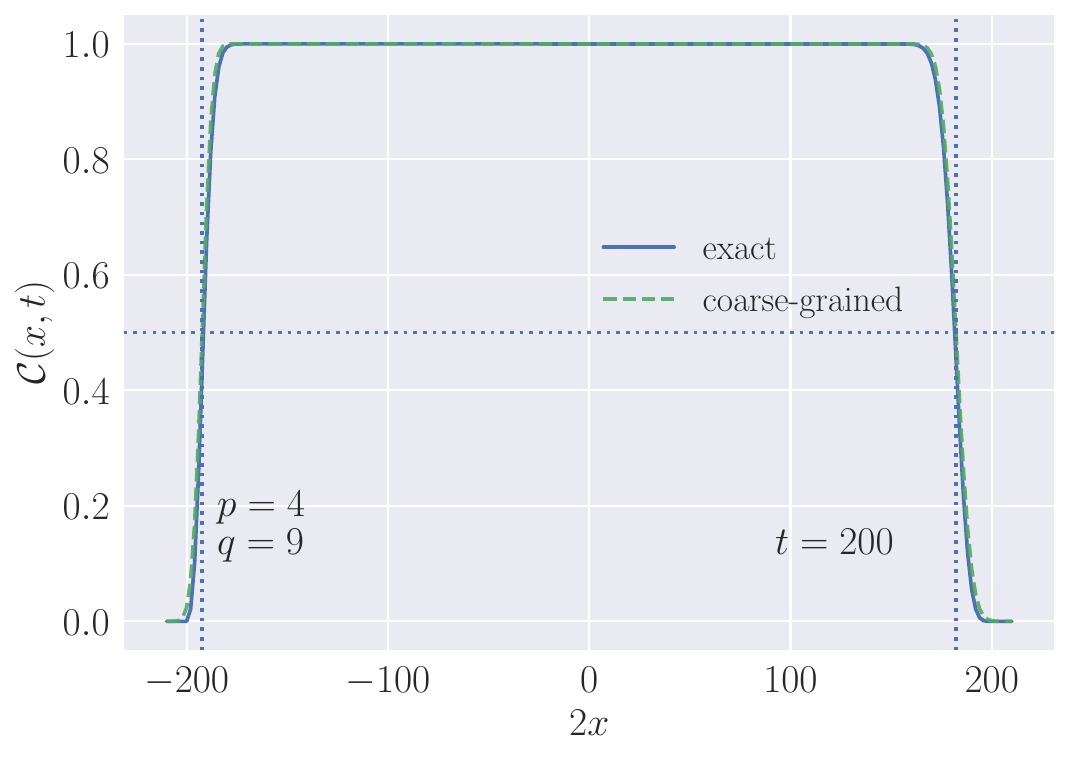}
	\includegraphics[scale=0.46]{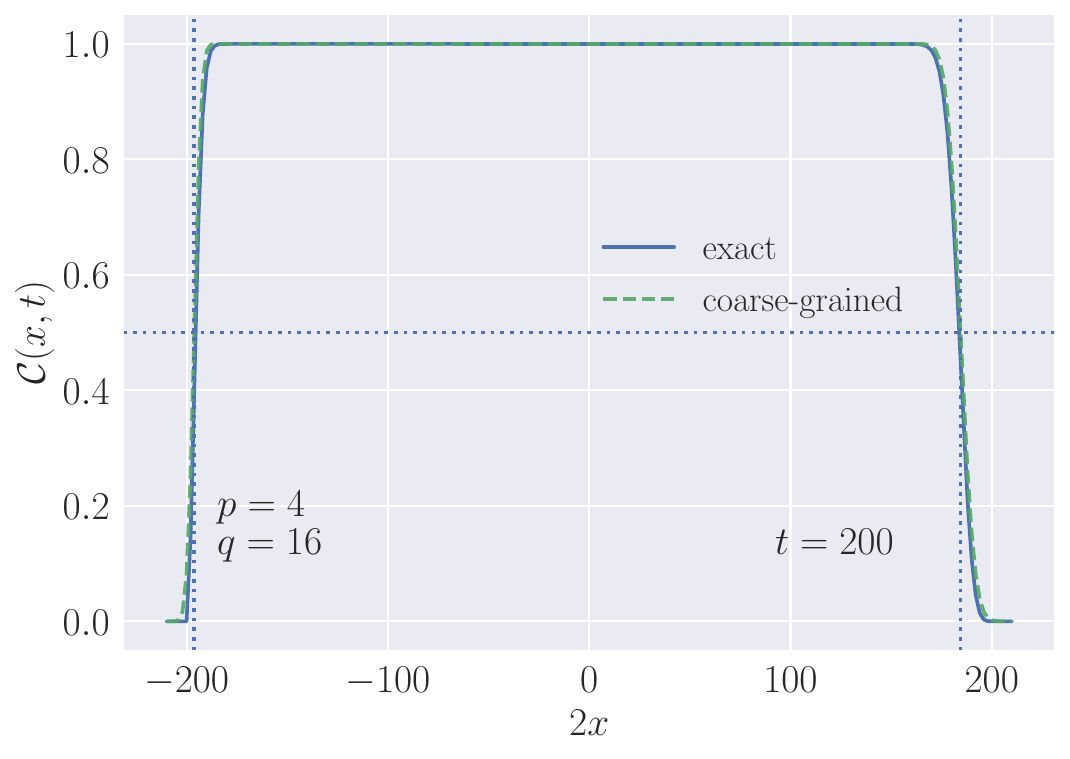}
	\caption{OTOC in Model A. Exact numerical results, obtained by evaluating~\eqref{eq:exact_OTOC}, are compared against the hydrodynamic description~\eqref{eq:asymptotic_otoc} for different values of the index. Vertical lines correspond to the left/right light-cone velocities. The curves are in excellent agreement and we verified that the difference decreases uniformly in $\mathbb{R}$ as $t$ increases.}
	\label{fig:comparison}
\end{figure}

Finally, let us consider the sums in Eqs.~\eqref{eq:g_L},~\eqref{eq:g_R}. They are dominated by the terms where $\exp[t \mathcal{H}(\alpha, r_p^{\pm 1})]d^{-2u}$ is largest, the others being negligibly small. This happens for values of $u$ that are of the same order of $t$, namely $u=\alpha t$.  The leading terms are thus once again determined via the saddle-point condition
\be
\frac{\partial}{\partial \alpha}\left[\mathcal{H}(\alpha, r_p^{\pm 1})-2\alpha \ln d\right]=0\,,
\ee
which yields
\be
\alpha^\ast_{\pm}=\frac{d^2 r_p^{\pm}+d^2+2 r_p^{\pm}}{\left(d^2+1\right) \left(d^2+r_p^{\pm}\right)}\,.
\ee
Recalling~\eqref{eq:exact_OTOC_large_t}, this gives us the left/right velocities
\begin{align}
	v_{L}&=1-\alpha_+=\frac{p^{2} q^{2}-q(p+q)+1}{p^{2} q^{2}-1}\,,\\
	v_{R}&=1-\alpha_-=\frac{p^{2} q^{2}-p(p+q)+1}{p^{2} q^{2}-1}\,.
\end{align}
Expanding up to the second order in $\alpha$ around the saddle points $\alpha^\ast_\pm$, we can also obtain the diffusive broadening of the light cones. Defining
\be
\mathcal{F}_\pm(\alpha)=-2\alpha \ln d+\mathcal{H}(\alpha,r_p^{\pm 1})
\ee
we get
\be
\mathcal{F}_\pm(\alpha)=\mathcal{F}_\pm(\alpha^\ast_\pm)-\frac{1}{2}(\alpha-\alpha^\ast_\pm)^2\Delta_\pm+O[(\alpha-\alpha^\ast_\pm)^3 ]\,,
\ee
where
\be
\Delta_\pm=\frac{\left(d^2+1\right)^2 \left(d^2+r_p^{\pm}\right)^2}{d^6 (r_p^{\pm}+1)+4 d^4 r_p^{\pm}+d^2 r_p^{\pm} (r_p^{\pm}+1)}\,.
\ee
Assuming that near $\alpha_\pm^\ast$ we can neglect the contributions coming from the subleading function $A(t,\alpha)$ (which varies slowly in $t$), we have that, in the region where $g_L(n,a)$ is not negligible, it is a discrete sum of  Gaussian weights. Therefore, we can take the continuum limit, and obtain the final result
\be\label{eq:asymptotic_otoc}
{\cal C}(x,t)\simeq \Phi\left(\frac{v_{L} t+x}{\sigma_L}\right) \Phi\left(\frac{v_{R} t-x}{\sigma_R}\right)\,,
\ee
where $\Phi(y)=\frac{1}{\sqrt{2 \pi}} \int_{-\infty}^{y} e^{-x^{2} / 2} \mathrm{~d} x $ and
\begin{align}
v_{L}&=\frac{p^{2} q^{2}-q(p+q)+1}{p^{2} q^{2}-1}\,,\qquad
v_{R}=\frac{p^{2} q^{2}-p(p+q)+1}{p^{2} q^{2}-1}\,,\label{eq:buttefly_velocities}\\
\sigma_L&=\frac{\sqrt{t} \sqrt{q \left[p^3 q^2+p^2 q \left(q^2-3\right)+p-q^3+q\right]}}{(p q-1) (p q+1)}\,, \qquad \sigma_R=\frac{\sqrt{t} \sqrt{p \left[p^2 q^3+\left(p^2-3\right) p q^2-p^3+p+q\right]}}{(p q-1) (p q+1)}\,.
\end{align}
We have tested numerically this result, see Fig.~\ref{fig:comparison} for an example. As an analytic check, we have verified that for $p=q=d$ we obtain the same formula for $\sigma$ in RUC obtained in~\cite{nahum2018operator}, after replacing $(x,t)$ with $(\frac{x}{2},\frac{t}{2})$. It is interesting to consider the the limit $p\to 1$. In this case, setting $q=d^2$ we find
\be
\lim_{p\to 1}\sigma_L=\lim_{p\to 1}\sigma_R=\frac{1}{\sqrt{2}} \frac{d\sqrt{2t}}{d^2+1}\,,
\ee
which is exactly the expression for a quantum circuit evolved up to time $\tilde{t}=t/2$. This is consistent, because a simple shift does not increase the operator support, and hence does not modify the broadening of the light cones. 

\subsection{Model B}
Similar calculations may be performed for Model B. In fact, in this case the analysis is significantly simpler, and here only report the final result.  Defining 
\begin{equation}\label{eq:pL_pR}
p_L=\frac{p^2-1}{p^2q^2-1},\;\;\;\;p_R=\frac{q^2-1}{p^2q^2-1},
\end{equation}
the OTOC for Model B reads
\begin{equation}
	\begin{split}
	{\cal C}(x,t)=\frac{p^2q^2}{p^2q^2-1} g(2t,t+x,p_L)g(2t,t-x,p_R) - \frac{1}{p^2q^2-1}g(2t,t+x-1,p_L)g(2t,t-x-1,p_R),
	\end{split}
\end{equation}
where
\begin{equation}
	g(n,m,p)\equiv\sum^m_{j=0} \binom{n}{j}  p^j(1-p)^{n-j}.
	\label{gnmp}
\end{equation}
In the hydrodynamic limit, ${\cal C}(x,t)$ is still of the form~\eqref{eq:asymptotic_otoc} where now
\begin{equation}
	\begin{split}
		v_L=1-2p_L=\frac{p^2q^2 - 2p^2+1 }{p^2q^2-1},& \;\;\;\;v_R=1-2p_R=\frac{p^2q^2 - 2q^2+1 }{p^2q^2-1},\\
		\sigma_L(t)=\sqrt{2p_L(1-p_L)t}=\frac{p\sqrt{2(p^2-1)(q^2-1)t}}{p^2q^2-1},& \;\;\;\;
		\sigma_R(t)=\sqrt{2p_R(1-p_R)t}=\frac{q\sqrt{2(p^2-1)(q^2-1)t}}{p^2q^2-1}.
	\end{split}
\end{equation}
Note that when $p=1$ ($q=1$), we have $v_L=-v_R=1$ ($v_R=-v_L=1$) and $\sigma_L(t)=\sigma_R(t)=0$, which is fully consistent with the left (right) shift. Furthermore, when $p=q$, we again reproduce the results in Ref.~\cite{nahum2018operator} after replacing $(x,t)$ with $(\frac{x}{2},\frac{t}{2})$.

\section{The operator entanglement and the tripartite information}
In this section we discuss different entanglement-related quantities. To this end, we define a doubled Hilbert space $\mathcal{H}=(\mathbb{C}^{d})^{\otimes 2L}\otimes (\mathbb{C}^{d})^{\otimes 2L}$ along with the maximally entangled state $\ket{I}=\ket{\phi_+}^{\otimes 2L}$, where $\ket{\phi_+}\equiv\sum^d_{n=1}\ket{n}\otimes\ket{n}/\sqrt{d}$ and $n$ runs over a basis of $\mathbb{C}^d$. 
This allows us to vectorize the unitary operator as the so-called Choi state $\ket{{\cal U}(t)}=({\cal U}(t)\otimes \openone)\ket{I}\in {\cal H}$, with the two Hilbert spaces naturally associated with its input and output degrees of freedom. All the entanglement-quantities can be obtained by calculating the entropies of the reduced states on properly chosen subsystems.

\subsection{Model A}
\begin{figure*}
	\begin{center}
		\includegraphics[scale=0.5]{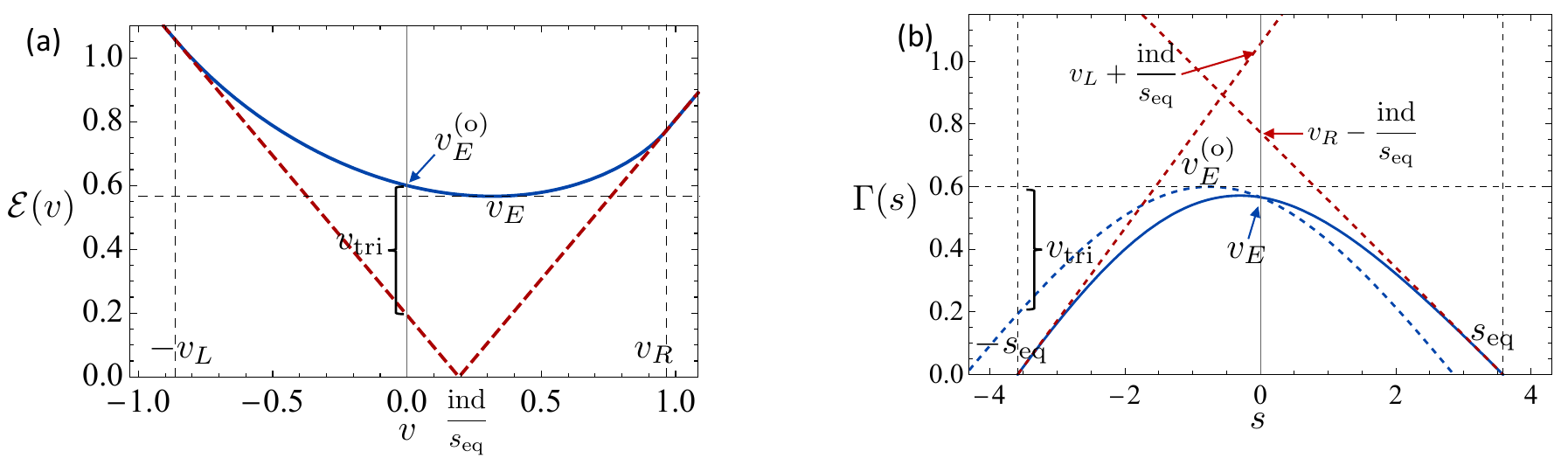}
	\end{center}
	\caption{(a) Entanglement line tension $\mathcal{E}(v)$ and (b) production $\Gamma(s)$ (in terms of R\'enyi-2 entropy) for Model A	with $d=6$ and $p=3$, $q=12$. The red dashed lines refer to the tangents at (a) $(v_{L,R},\mathcal{E}(\mp v_{L,R})=v_{L,R}\pm \ind/s_{\rm eq})$; (b) $(\pm s_{\rm eq},0)$. The blue dashed line in (b) corresponds to $\Gamma(s)-(\ind/s_{\rm eq})(s/s_{\rm eq})$, whose maximum gives $v^{(\rm o)}_E$.}
	\label{fig:SM_membrane}
\end{figure*}

Considering a bipartition of the system with boundary at site $x$ and $y$ in the input and output, respectively, our goal is to compute the bipartite entanglement  $S^{(\mathrm{o})}(x,y, t)$ of $\ket{{\cal U}(t)}$ for an infinite system ($L\to\infty$). Exploiting translation symmetry, we have $S^{(\mathrm{o})}(x,y, t)=S^{(\mathrm{o})}(x-y, t)$. As mentioned, although computing R\'enyi-$n$ entropies exactly is very challenging, our model makes it possible to obtain an analytic expression for the average of the purity, i.e. $\overline{e^{-S^{(\rm o)}_2(x,t)}}$. Indeed, this quantity may be expressed in terms of two copies of ${\cal U}(t)\otimes {\cal U}^\ast(t) $, and can be computed exploiting the very same mapping to an Isinig partition function in $2$D used for the OTOCs~\cite{nahum2017quantum,nahum2018operator}. As before, we omit the details of this procedure which follow closely those for RUC. We obtain the following exact expression valid at finite $x$ and $t$
\begin{equation}
	\overline{e^{-S^{(\rm o)}_2(x,t)}}= \left(\frac{d}{d^2+1}\right)^t \sum^{2t}_{u=0}\sum^u_{a=0} 
	\binom{t}{a} 	\binom{t}{u-a} \left[\frac{q(p^2-1)}{p^2q^2-1}\right]^a\left[\frac{p(q^2-1)}{p^2q^2-1}\right]^{t-a} (pq)^{-|u-t-x|}\,,
	\label{So2sum}
\end{equation}
which may be simplified as
\begin{equation}\label{eq:purity_analytic}
\overline{e^{-S^{(\rm o)}_2(x,t)}} =\left[\frac{dp(q^2-1)}{(d^2+1)(p^2q^2-1)}\right]^t \sum^{2t}_{u=0} f(t,u,r_p^{-1}) (pq)^{-|u-t-x|}\,,
\end{equation}
where $f(t,x,r)$ and $r_p$  are defined in~\eqref{ftxr}, ~\eqref{eq:rp} respectively, and we used the identity
\begin{equation}
	f(t,x,r)=r^t f (t,2t-x,r^{-1})\,.
\end{equation}
In order to extract information on the line tension, we need to analyze the hydrodynamic limit of~\eqref{eq:purity_analytic}. This follows once again from~\eqref{fasym}. Indeed, we first note that, within the light cone, we have the following leading behavior
\begin{equation}
	\overline{e^{-S^{(\rm o)}_2(x,t)}} \simeq \left[\frac{dp(q^2-1)}{(d^2+1)(p^2q^2-1)}\right]^t f(t,t+x,r_p^{-1}) 
	= \left[\frac{d^2 \sqrt{(p^2-1)(q^2-1)}}{(d^2+1)(p^2q^2-1)\sqrt{r_p}}\right]^t f(t,t-x,r_p). 
\end{equation}
Therefore, the R\'enyi-2 entanglement line tension
\begin{equation}
	\mathcal{E}_2(v) \equiv -\lim_{t\to\infty} \frac{\ln \overline{e^{-S^{(\rm o)}_2(vt,t)}}}{t s_{\rm eq}}
	=\log_{pq}\frac{(pq+1)^2(pq-1)\sqrt{r_p}}{pq\sqrt{(p^2-1)(q^2-1)}} -\lim_{t\to\infty} \frac{\ln f(t,(1-v)t,r_p)}{t\log(pq)}
\end{equation}
becomes
\begin{equation}
	\mathcal{E}_2(v) \!=\! \log_{pq}\!\frac{(pq+1)^2(pq-1)}{pq \sqrt{(p^2-1)(q^2-1)}} -\! v\log_{pq}\!\frac{\sqrt{(1-r_p)^2v^2+4r_p}-(1+r_p)v}{2(1+v)} -\!\log_{pq}\! \frac{1 + r_p + \sqrt{(1-r_p)^2v^2+4r_p}}{(1-v^2)\sqrt{r_p}}\,.
	\label{SM:E2v}
\end{equation}
The main qualitative features of  $\mathcal{E}_2(v)$ can be seen from Fig.~\ref{fig:SM_membrane}, where we have plotted it for given values of $d$, $p$ and $q$.

Eq.~\eqref{SM:E2v} allows us to extract several quantities and check explicitly a few aspects of the generalized EMT. First al all, the value of $\mathcal{E}_2(v)$ in $v=0$ defines the operator R\'enyi-2 entanglement velocity
\begin{equation}
	v^{(\rm o)}_E  = \mathcal{E}_2(0) = \log_{pq}\frac{(pq+1)^2(pq-1)}{\sqrt{pq}(\sqrt{q(p^2-1)} + \sqrt{p(q^2-1)})^2}\,.
	\label{voE}
\end{equation}
Next, its minimum value determines the entanglement speed $v_E$~\cite{nahum2017quantum,jonay2018coarse}. To compute it, we need to solve $\mathcal{E}_2'(v)=0$. While this is a complicated transcendental equation, we can eliminate the intractable logarithm term by making the ansatz 
\be\label{eq:ansatz}
\sqrt{(1-r_p)^2v^2+4r_p}-(1+r_p)v = 2(1+v)\,,
\ee
which leads to the solution
\begin{equation}
	v_m=\frac{r_p -1}{2(r_p+1)}\,.
	\label{SM:vmin}
\end{equation}
It is immediate to verify that $v_m$ found in this way indeed satisfies~\eqref{eq:ansatz}. Plugging~\eqref{SM:vmin} into~\eqref{SM:E2v} we obtain the entanglement velocity
\begin{equation}
	v_E= \log_{pq} \frac{(pq+1)^2}{2\sqrt{pq}(p+q)}.
	\label{vp}
\end{equation}

We can check explicitly that this is the growth rate of the state R\'enyi-$2$ entropy after a quench. Indeed, the latter is readily computed for an initial product state, and reads
\begin{equation}
	\overline{e^{-S_2(t)}}= \left[\frac{dp(q^2-1)}{(d^2+1)(p^2q^2-1)}\right]^t \sum^{2t}_{u=0} f(t,u,r_p^{-1}) = 
	\left[\frac{2dp(q^2-1)(1+r^{-1}_p)}{(d^2+1)(p^2q^2-1)}\right]^t = \left[ \frac{2d(p+q)}{(d^2+1)(pq+1)}\right]^t\,,
\end{equation}
from which we see that  $-\lim_{t\to \infty} \ln \overline{e^{-S_2(t)}}/t s_{\rm eq}$ equals $v_E$ in~\eqref{vp}. Finally, we can also easily check that the butterfly velocities~\eqref{eq:buttefly_velocities} satisfy the identities
\begin{align}
	\mathcal{E}_2(v_R)&=v_R - \frac{\ind}{s_{\rm eq}},\;\;\;\;\mathcal{E}_2(-v_L)=v_L + \frac{\ind}{s_{\rm eq}}\,,\label{eq:identities_vel_1}\\
	\mathcal{E}'_2(v_R)&=-\mathcal{E}'_2(-v_L)=1\,.\label{eq:identities_vel_2}
\end{align}

For $v> v_R$ or $v< -v_L$, a direct computation also  shows
\begin{equation}
	\mathcal{E}_2 (v) = \left|v - \frac{\ind}{s_{\rm eq}} \right|\,.
	\label{eq:gv}
\end{equation}

Eq.~\eqref{eq:gv} is a natural generalization of $\mathcal{E} (v)=|v|$, $\forall |v|>v_B$ for spatial-reflection symmetric quantum circuits \cite{jonay2018coarse}. Importantly, although it was derived for random QCA, it holds more generally, and we can show that, for sufficiently large $v$,  it is valid for arbitrary entanglement R\'enyi entropy and any QCA, even without spacetime disorder. To see this, we consider the case of $v>0$. Suppose that $vt$ is larger than the range of the QCA at time $t$, which has index $\ind t$ due to its additivity, we can take a specific bilayer form \cite{gross2012index} such that a local operator (LO) with respect to the entanglement cut can transform the operator into an array of virtual identities: 
\begin{equation}
	\begin{tikzpicture}
		\fill[red!10!white] (3.5,-0.8) -- (1,-0.8) -- (1,0) -- (1.5,0) -- (1.5,0.8) -- (3.5,0.8) -- cycle;
		\foreach \x in {0,...,3}
		{
			\draw[ultra thick] (\x-0.75,-0.1) -- (\x-0.75,0.1);
			\draw[ultra thick,dotted] (\x-0.25,-0.1) -- (\x-0.25,0.1);
			\draw[thick,fill=orange!20!white] (\x-0.4,0.1) rectangle (\x+0.4,0.5);
			\draw[thick,fill=yellow!20!white] (\x-0.9,-0.1) rectangle (\x-0.1,-0.5);
		}
		\foreach \x in {-1,...,7}
		{
			\draw[thick] (0.5*\x-0.25,-0.5) -- (0.5*\x-0.25,-0.7);
			\draw[thick] (0.5*\x-0.25,0.5) -- (0.5*\x-0.25,0.7);
		}
		\draw[ultra thick] (3.25,-0.1) -- (3.25,0.1);
		\fill[orange!20!white] (-1,0.1) rectangle (-0.6,0.5);
		\fill[yellow!20!white] (3.5,-0.1) rectangle (3.1,-0.5);
		\draw[thick] (-1,0.1) -- (-0.6,0.1) -- (-0.6,0.5) -- (-1,0.5);
		\draw[thick] (3.5,-0.1) -- (3.1,-0.1) -- (3.1,-0.5) -- (3.5,-0.5);
		\draw[dashed] (1,-0.8) -- (1,0) -- (1.5,0) -- (1.5,0.8);
		\draw[thick] (1.1,0.7) .. controls (1.1,0.9) and (1.25,0.7) .. (1.25,0.9) (1.4,0.7) .. controls (1.4,0.9) and (1.25,0.7) .. (1.25,0.9);
		\Text[x=1.25,y=1.1]{$vt$}
		\begin{scope}[>=latex] 
			\draw[->] (4,0) -- (5,0);
		\end{scope}
		\Text[x=4.5,y=0.25]{LO}
	\end{tikzpicture}
	\;\;\;\;\;\;
	\begin{tikzpicture}
		\fill[red!10!white] (3.5,-0.8) -- (1,-0.8) -- (1,0) -- (1.5,0) -- (1.5,0.8) -- (3.5,0.8) -- cycle;
		\foreach \x in {0,...,3}
		{
			\draw[ultra thick] (\x-0.75,-0.4) -- (\x-0.75,0.4);
			\draw[ultra thick,dotted] (\x-0.25,-0.4) -- (\x-0.25,0.4);
		}
		\draw[ultra thick] (3.25,-0.4) -- (3.25,0.4);
		\draw[dashed] (1,-0.8) -- (1,0) -- (1.5,0) -- (1.5,0.8);
	\end{tikzpicture}
\end{equation}
Obviously, the operator entanglement, irrespective of its R\'enyi order, is the logarithm of the dimension of the virtual Hilbert space across the cut, which reads  
\begin{equation}
	S^{(\rm o)}_\alpha(vt,t)= \ln (s_{\rm eq}^{vt} e^{-\ind t}) = s_{\rm eq} vt - \ind t\,,
\end{equation}
Similarly, for $v<0$ we have
\begin{equation}
	S^{(\rm o)}_\alpha(vt,t)= \ln (s_{\rm eq}^{-vt} e^{\ind t}) = - s_{\rm eq} vt + \ind t\,,
\end{equation}
as anticipated.

As discussed in the main text, given the line tension ${\cal E}_2(v)$ we can define the entanglement production rate as
\begin{equation}
	\Gamma_2 (s) = \min_v \left[\mathcal{E}_2(v) - \frac{vs}{s_{\rm eq}}\right] + \frac{\ind}{s_{\rm eq}}\frac{s}{s_{\rm eq}}\,.
	\label{SM:GE}
\end{equation}
From a straightforward calculation, we obtain
\begin{equation}\label{eq:gamma_2}
	\Gamma_2 (s) = {\cal E}_2(\vbkg_s) - \frac{\vbkg_ss}{s_{\rm eq}} + \frac{\ind}{s_{\rm eq}}\frac{s}{s_{\rm eq}}\,,
\end{equation}
where 
\be
\vbkg_s=\frac{r_p e^{2 s}-1}{\left(e^s+1\right) \left(r_p e^s+1\right)}\,,
\ee
and $r_p$ is given in~\eqref{eq:rp}. The main qualitative features of  $	\Gamma_2 (s) $ can be see from Fig.~\ref{fig:SM_membrane}, where we have plotted it for given values of $d$, $p$ and $q$. Note in particular, that it satisfies the following relations
\begin{equation}
	- s_{\rm eq} \Gamma' (s_{\rm eq}) = v_R - \frac{\ind}{s_{\rm eq}},\;\;\;\;
	s_{\rm eq} \Gamma' (-s_{\rm eq}) = v_L + \frac{\ind}{s_{\rm eq}}.
\end{equation}

As anticipated, our model makes it possible to also compute another useful measure of scrambling, the so-called tripartite mutual information~\cite{hosur2016chaos}, see also~\cite{sunderhauf2019quantum,schnaack2019tripartite,bertini2020scrambling,kudler2021information}. We recall that, given two bipartitions $A\cup B$ and $C\cup D$ with $A=(-\infty,0)$, $B=(0,\infty)$, $C=(-\infty,x)$, $D=(x,\infty)$ for the input and output, respectively, the tripartite information is given by $I^{A:C:D}(x,t)\equiv I^{A:C}(x,t) + I^{A:D}(x,t) - I^{A:CD}(x,t)$. Using the unitarity of $\mathcal{U}(t)$, we can simplify  $I^{A:C}(x,t)$ into $ S_{\max} - S^{AD}(x,t) -  S^{AC}(x,t)$, where $S_{\max}= 2L \ln d$. Taking the R\'enyi-$2$ version,  $S^{AC}(x,t)$ is nothing but $S^{(\rm o)}_2(x,t)$, whose exponentiated average has been obtained previously. Therefore, all we have to do is to compute $S_{\max} - S^{AD}(x,t)$.

While both $S_{\max}$ and $S^{AD}(x,t)$ diverge in the thermodynamic limit, their difference remains finite. Moreover, we can again make use of the map to the Ising model, obtaining  
\begin{equation}
\overline{e^{S_{\max} - S^{AD}_2(x,t)}} =  \left(\frac{d}{d^2+1}\right)^t \sum^{2t}_{u=0}\sum^u_{a=0} 
	\binom{t}{a} 	\binom{t}{u-a} \left[\frac{q(p^2-1)}{p^2q^2-1}\right]^a\left[\frac{p(q^2-1)}{p^2q^2-1}\right]^{t-a} (pq)^{|u-t-x|}\,.
\end{equation}
After some straightforward calculations, we can relate the above quantity to $g_{L,R}$ in Eqs.~(\ref{eq:g_L}) and (\ref{eq:g_R}):
\begin{equation}
\overline{e^{S_{\max} - S^{AD}_2(x,t)}} \simeq \left(\frac{q}{d}\right)^t d^{2x}  g_R(2t-1,t+x) + \left(\frac{p}{d}\right)^t d^{-2x}  g_L(2t-1,t-x).  
\end{equation}
Since $g_{L,R}$ (almost) saturates $1$ inside the light cone, the dynamics of the R\'enyi-2 tripartite information should be governed by
\begin{equation}
\overline{ e^{I^{A:C:D}_2(0,t)}}\simeq \overline{e^{S_{\max} - S^{AD}_2(0,t)}} \overline{e^{-S^{\rm o}_2(t)}} \simeq \max\left\{\left(\frac{p}{d}\right)^t,\left(\frac{q}{d}\right)^t\right\} d^{ 2v^{(\rm o)}_E t},
\end{equation}
where $v^{(\rm o)}_E$ is given in Eq.~(\ref{voE}). Noting that $|\ind|=\ln\max\{p/d,q/d\}$, we have
\begin{equation}
v_{\rm tri}\equiv-\lim_{t\to\infty}\frac{\overline{I^{A:C:D}_2(0,t)}}{t s_{\rm eq}}= v^{(\rm o)}_E - \frac{|\ind| }{s_{\rm eq}}\,.
\label{vtri}
\end{equation}

\begin{figure*}
	\begin{center}
		\includegraphics[width=12cm, clip]{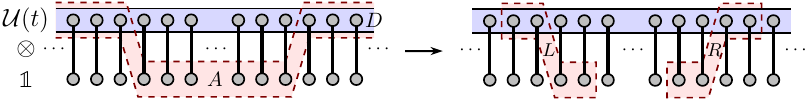}
	\end{center}
	\caption{For a QCA $\mathcal{U}(t)$, $S_{\max} - S^{AD}$ of its Choi state $|\mathcal{U}(t)\rangle$ is equal to the sum of $S^L_{\max} - S^L$ and $S^R_{\max} - S^R$, since the reduced state on $AD$ is factorized as $\rho_{AD}= \rho_L\otimes \rho_R\otimes \openone_{AD\backslash LR}/\Tr \openone_{AD\backslash LR}$ \cite{gong2021topological}, provided that $L$ and $R$ are large enough.}
	\label{fig:SLR}
\end{figure*}

We conjecture that Eq.~(\ref{vtri}) is valid universally for generic anomalous dynamics with translation invariance on the ensemble level. A heuristic argument is as follows: for a while let us take the periodic boundary condition so that $S_{\max} - S^{AD}$ can be decomposed into two terms arising from the left and right boundaries (cf. Fig.~\ref{fig:SLR}): 
\begin{equation}
S_{\max} - S^{AD} = S^L_{\max} - S^L + S^R_{\max} - S^R. 
\end{equation}
Suppose that the segments across the left and right boundary are taken to be of the same length and using the translation invariance of the random QCA on the ensemble level, we know from the entropy formula of the index \cite{gong2021topological} that
\begin{equation}
|S^L - S^R|\simeq 2|\ind| t.
\end{equation}
As long as we can justify 
\begin{equation}\label{eq:to_justify}
\max\{S^L,S^R\}\simeq S^{L,R}_{\max}\,,
\end{equation}
we immediately obtain $S_{\max} - S^{AD}\simeq 2|\ind| t$, implying $ t^{-1}(S_{\max} - S^{AD})\simeq |\ind|  $ for the half-infinite entanglement cut. For model A, ~\eqref{eq:to_justify} can be established for the R\'enyi-$2$ entropy and so it should hold for arbitrary R\'enyi (including von Neumann) entropies, under the assumption of small fluctuations. Therefore, since it holds for Haar-random QCA, we expect it to hold also for sufficiently chaotic dynamics. As a consistency check, we will see that Eq.~\eqref{vtri} holds true also for Model B.

\subsection{Model B}
\begin{figure*}
	\begin{center}
		\includegraphics[scale=0.55]{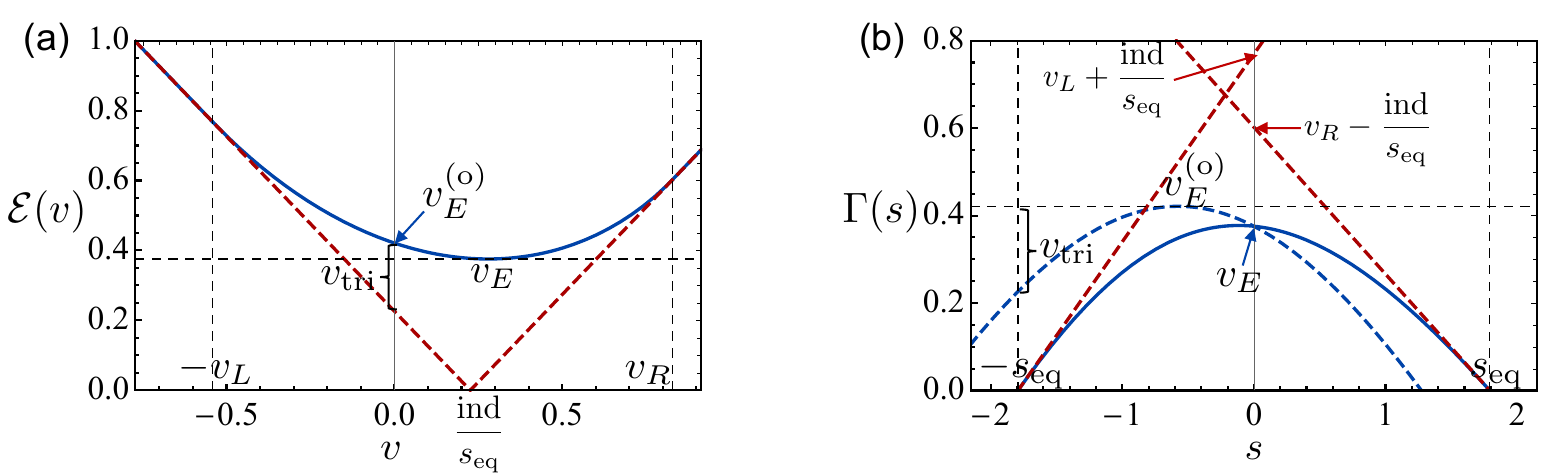}
	\end{center}
	\caption{(a) Entanglement line tension $\mathcal{E}(v)$ and (b) production $\Gamma(s)$ (in terms of R\'enyi-2 entropy) for Model B with $p=3$ and $q=2$. The red and blue dashed lines share the same meaning as those in Fig.~\ref{fig:SM_membrane}.}
	\label{fig:SM_membrane2}
\end{figure*}

For completeness, let us also briefly discuss  how the previous results read for Model B.  Since the calculations are analogous, and in fact significantly simpler, we omit the derivation. First, for the R\'enyi-$2$ operator entanglement we find 
\begin{equation}
	\begin{split}
		\overline{e^{-S^{(\rm o)}_2(x,t)}}&= \frac{p^{t-x}}{q^{t+x}} g(2t,t+x-1,1-p_R) + \frac{q^{t+x}}{p^{t-x}}g(2t,t-x,1-p_L) \\
		&\simeq {\rm C}^{t+x}_{2t} \left\{\left[\frac{p}{q}(1-p_R)p_R\right]^t\left(\frac{1-p_R}{p_R}\right)^x (pq)^{-x} + \left[\frac{q}{p}(1-p_L)p_L\right]^t\left(\frac{p_L}{1-p_L}\right)^x (pq)^x\right\},
	\end{split}
\end{equation}
where $p_L$, $p_R$ are defined in~\eqref{eq:pL_pR}, while $g(n,m,p)$ is given in~\eqref{gnmp}. This allows us to obtain the line tension
\begin{equation}
	\mathcal{E}_2(v)= \lim_{t\to \infty}\frac{S^{(\rm o)}_2(vt,t)}{\ln (pq) t}
	=\log_{pq}\frac{(p^2q^2-1)^2}{pq(p^2-1)(q^2-1)} + (1+v)\log_{pq}\frac{1+v}{2} + (1-v)\log_{pq}\frac{1-v}{2} + v\log_{pq}\frac{p(q^2-1)}{q(p^2-1)}\,.
	\label{Ev}
\end{equation}
From~\eqref{Ev}, we readily read off the values of the operator ($v_E^{(o)}$) and state ($v_E$) entanglement velocities
\begin{align}
	v^{(\rm o)}_E &= \mathcal{E}_2(0) = \log_{pq}\frac{(p^2q^2-1)^2}{4pq(p^2-1)(q^2-1)}\,,\label{eq:model_B_voe}\\
	v_E &= \min_v \mathcal{E}_2(v) = 2\log_{pq}\frac{pq+1}{p+q}\,,\label{eq:modelB_vE}
\end{align}
where the minimum is attained at $v_m=\frac{(p-q)(pq+1)}{(p+q)(pq-1)}$. Once again, it is straightforward to check that $v_E$ coincides with the growth rate of  the R\'enyi-$2$ entropy after a quench from an initial product state, and that the  identities~\eqref{eq:identities_vel_1}, ~\eqref{eq:identities_vel_2} hold. Finally, from the previous results, we also obtain an explicit expression for the entanglement production rate~\eqref{eq:gamma_2}
\begin{equation}
	\Gamma_2 (s) = 2\log_{pq}\frac{pq+1}{p+q} - 2\log_{pq} \frac{p(q^2-1) e^{-\frac{1}{2}(1+\frac{\ind}{s_{\rm eq}})s} + q(p^2-1) e^{\frac{1}{2}(1-\frac{\ind}{s_{\rm eq}})s}}{(pq-1)(p+q)}\,,
\end{equation}
from which one can simply verify the validity of Eqs.~\eqref{SM:GE}. See Fig.~\ref{fig:SM_membrane2} for some typical profiles of $\mathcal{E}_2(v)$ and $\Gamma_2(s)$.

Next, using sing the same approximation employed for Model A, we can also compute the (averaged) tripartite information. In particular, we have
\begin{equation}
	\begin{split}
		e^{\overline{I^{A:C:D}_2(x,t)}}&\simeq \tilde{g}_L(1-\tilde{g}_R)+{\tilde g}_R(1-{\tilde g}_L)+\frac{p^{2t-2x}}{q^{2t+2x}}{\tilde g}_R(1-{\tilde g}_R)+\frac{q^{2t+2x}}{p^{2t-2x}}{\tilde g}_L(1-{\tilde g}_L)\,,
	\end{split}
\end{equation}
where we defined
\begin{equation}
{\tilde g_L}=g(2t,t+x-1,p_L)\,,\qquad {\tilde g_R}=g(2t,t-x,p_R)\,,
\end{equation}
with $p_L$, $p_R$ and $g(n,m,p)$ given in~\eqref{eq:pL_pR} and~\eqref{gnmp} respectively. We see that when the spacetime coordinate is outside the (strict) light cone, i.e., $x>t$ (or $x<-t$), we have $e^{\overline{I^{A:C:D}_2(x,t)}}\simeq1$, consistent with the intuition that the quantum information cannot propagate faster than ``light''.  Furthermore, for the translation QCA with $q=1$ ($p=1$), we have again $e^{\overline{I^{A:C:D}_2(x,t)}}\simeq1$. This result is consistent with the observation that no (quantum) scrambling takes place in swap (permutation) circuits~\cite{hosur2016chaos}. Finally, if we take $p=q$, we reproduce the results of Ref.~\cite{bertini2020scrambling} for RUC, as we should. Taking now $x=0$, we obtain
\begin{equation}
	e^{\overline{I^{A:C:D}_2(0,t)}}\simeq e^{[|\ind| - v^{(\rm o)}_E\ln(pq)]t},
\end{equation}
where $v^{(o)}_E$ is given in Eq.~\eqref{eq:model_B_voe}, implying the validity of Eq.~(\ref{vtri}). 

\end{document}